\documentclass[aps,prd,amsmath,floats,twocolumn,floatfix,nofootinbib,showpacs,superscriptaddress]{revtex4}
\usepackage{graphicx} 
\usepackage{bm}
\usepackage{mathrsfs}

\usepackage[usenames]{color}
\usepackage{ulem}
\newcommand{\Mirr}{M_{\text{irr}}}

\newcommand{\Lapse}{\alpha}
\newcommand{\ScriPlus}{\mathscr{I}^+}

\newcommand{\Rrescale}{\eta}

\newcommand{\rT}{r_{\rm T}}
\newcommand{\KT}{K_{\rm T}}
\newcommand{\CT}{C_{\rm T}}
\newcommand{\MT}{M_{\rm T}} 

\newcommand{\TrKW}{K}
\newcommand{\KW}{K}
\newcommand{\AW}{A}

\newcommand{\Caltech}{\affiliation{Theoretical Astrophysics,
    California Institute of Technology, Pasadena, CA 91125}}

\newcommand{\Texas}{\affiliation{Center for Relativity, University of Texas at
Austin, Austin, Texas 78712}}

\newcommand{\Washington}{\affiliation{Physics Department, University of Washington, Seattle, WA 98195}}

\newcommand{\CITA}{
        \affiliation{Canadian Institute for Theoretical Astrophysics,
        University~of~Toronto, Toronto, Ontario M5S 3H8, Canada}}

\begin{document}

\title{Black hole initial data on hyperboloidal slices}

\author{Luisa T. Buchman}\Caltech\Texas
\author{Harald P. Pfeiffer} \Caltech\CITA
\author{James M. Bardeen} \Washington

\date{\today}

\begin{abstract}
  We generalize Bowen-York black hole initial data to hyperboloidal
  constant mean curvature slices which extend to future null
  infinity. We solve this initial value problem numerically for
  several cases, including unequal mass binary black holes with spins
  and boosts. The singularity at null infinity in the Hamiltonian
  constraint associated with a constant mean curvature hypersurface
  does not pose any particular difficulties. The inner boundaries of
  our slices are minimal surfaces. Trumpet configurations are explored
  both analytically and numerically.
\end{abstract}


\maketitle

% For figures, the makefile is set up to automatically
% convert any .agr file to an .eps file.
%includefile[width=3in]{FilenameWithoutDotEpsSuffix}

%%%%%%%%%%%%%%%%%%%%%%%%%%
\section{Introduction}
%%%%%%%%%%%%%%%%%%%%%%%%%%

For asymptotically flat spacetimes, gravitational radiation is
well-defined only at future null infinity ($\ScriPlus$)
\cite{Bondi1962,Sachs1962,Sachs1964}. Consequently, numerical
relativists extracting gravitational waves from binary black hole
simulations ideally would like to include $\ScriPlus$ in their
computational domains, so that the Bondi news
function~\cite{Bondi1962,Stewart1989} (which contains the
gravitational wave information) can be computed. Additionally,
extending the simulation to $\ScriPlus$ would make it unnecessary to
deal with gravitational wave extraction at a finite distance, or with
artificial outer boundaries on a truncated domain, two very
complicated aspects of black hole simulations (see,
e.g.~\cite{Nerozzi2005,Campanelli2006,Lehner2007,
  Calabrese2001,Rinne2006,Buchman2006,Rinne2007,Ruiz2007,
  Rinne2008b,Seiler2008}).

Null infinity can be included in the computational domain via a
compactified radial coordinate on the null hypersurfaces of the
characteristic initial value problem~\cite{Winicour2009} , but the
null hypersurfaces are subject to caustic singularities, particularly
in the strong fields around black holes. The caustics can be avoided
by Cauchy-characteristic matching, but an appealing alternative is
solving the Cauchy problem on conformally compactified hyperboloidal
spacelike slices. These behave like conventional $3+1$ slicing in the
vicinity of the sources, but smoothly become asymptotically null as
they approach null infinity at a finite coordinate distance
\cite{Stewart1982,Friedrich1983,FrauendienerLRR}.
Friedrich~\cite{Friedrich1983} derived a system of symmetric
hyperbolic evolution equations based on the Bianchi identities for the
conformal Weyl tensor which are regular at null infinity on
hyperboloidal hypersurfaces provided certain smoothness conditions are
satisfied. This system has been used with some, but limited, success
in numerical calculations (see, e.g. \cite{Huebner1999}). More
recently, evolution schemes have been proposed which directly evolve
the conformal metric and the conformal factor through the conformally
compactified the Einstein equations on hyperboloidal
hypersurfaces~\cite{Zenginoglu2008,Moncrief2009}. They are not
manifestly regular at $\ScriPlus$, but in \cite{Moncrief2009},
Moncrief and Rinne derive regularity conditions based on the
constraint equations which deal successfully with the singularity of
the conformal factor at $\ScriPlus$ and can be imposed in a numerical
implementation.

Numerical evolution schemes on hyperboloidal hypersurfaces extending
to $\ScriPlus$ require initial data. The initial value problem on
asymptotically null hyperboloidal constant mean curvature (CMC) slices
turns out to be remarkably similar to the corresponding problem on
asymptotically flat {\em zero} mean curvature slices, and can be
attacked similarly to the well-known conformally flat
Bowen-York~\cite{Bowen-York:1980} initial data. (The Bowen-York
initial data presented in~\cite{Bowen-York:1980} also forms the basis
of puncture data~\cite{Brandt1997}\footnote{Puncture data differs from
  inversion symmetric data in the handling of the black hole
  singularities; we do not discuss the puncture treatment of
  singularities on CMC slices, although all our results concerning the
  constraint equations will carry over to such a
  treatment.}). Therefore, in this paper, we shall refer to the
initial data constructed as {\em hyperboloidal Bowen-York data}.

There are four main points to this paper: (i) to lay down the
formalism for constructing hyperboloidal Bowen-York black hole initial
data on CMC slices containing one or more black holes with arbitrary
masses, spins and boosts, (ii) to give rules for choosing the various
free parameters entering the formalism, (iii) to understand the
physical meaning of the free parameters and (iv) to discuss the
physical interpretation of the constructed solutions, which is
different for CMC slices than for traditional maximal slices.

Part of the initial data construction is the numerical solution of an
elliptic equation arising from the Hamiltonian constraint. On
hyperboloidal slices, this equation is formally singular at the outer
boundary $\ScriPlus$. Nevertheless, the employed spectral elliptic
solver~\cite{Pfeiffer2003} does not exhibit any problems while
computing the solution.

Our initial data is formulated with a singularity-avoiding minimal
surface boundary condition, on an Einstein-Rosen bridge connecting two
asymptotically flat ends. In the limit in which the conformal radius
approaches zero, a ``trumpet'' configuration is formed in which the
Einstein-Rosen bridge becomes infinitely long in proper distance.
Hyperboloidal slicings that contain trumpets have been examined in
Refs.~\cite{Zenginoglu2008b,Hannam2009}. Trumpet initial data is of
interest to numerical relativists using the moving puncture
approach~\cite{Campanelli2006a,Baker2006a} because puncture initial
data evolve quickly toward a trumpet
configuration~\cite{Hannam2006,Hannam2007d,Hannam2008}.

The organization of this paper is as follows: Section~\ref{Sec:2}
presents and analyzes the formalism for constructing hyperboloidal
Bowen-York initial data. Specifically, Sec.~\ref{Sec:IVP} presents the
initial value formalism on CMC slices and Sec.~\ref{Sec:Schwarzschild}
gives the particulars in the special case of a Schwarzschild black
hole. Sec.~\ref{sec:MinimalSurfacesAndTrumpets} discusses the
parameter space yielding a minimal surface at the inner boundary of a
Schwarzschild black hole in CMC slicing. Sec.~\ref{Sec:SchwarzIVP}
presents the construction of initial data for single black holes, with
and without Bowen-York spin and boost parameters. The physical
interpretation of the Bowen-York parameters is also discussed. In
Sec.~\ref{Sec:MultiBhIVP}, the methods presented for single black
holes are generalized to multiple black holes. Sec.~\ref{Sec:Trumpet}
details the solution to the Hamiltonian constraint equation when the
inner boundary is a trumpet. Sec.~\ref{Sec:3} presents our numerical
solutions for hyperboloidal Bowen-York initial data. Specifically,
Secs.~\ref{Sec:NumResults:Schwarz},~\ref{Sec:NumResults:SingleSpin}
and~\ref{Sec:NumResults:SingleBoost} give solutions for single black
holes that are spherically symmetric, spinning, or boosted,
respectively. Sec.~\ref{Sec:NumResults:BBH} presents numerical results
for two unequal mass black holes with arbitrarily oriented spins and
boosts. Finally, Sec.~\ref{Sec:Discussion} discusses our results. An
Appendix is included, in which conditions are derived for inversion
symmetry on both CMC and maximal slices.

%%%%%%%%%%%%%%%%%%%%%%%%%%
\section{Analytics}
\label{Sec:2}
%%%%%%%%%%%%%%%%%%%%%%%%%%
\subsection{Initial value formalism on CMC slices}
\label{Sec:IVP}

With a standard 3+1 decomposition~\cite{ADM,york79}, the spacetime metric 
is written as
\begin{equation}
ds^2 \,=\, -\alpha^2 \, dt^2 \,+\, g_{ij}\,(dx^i +
\beta^i\,dt)\,(dx^j+\beta^j\,dt),
\end{equation}
where $g_{ij}$ represents the induced metric on the $t=\mbox{const}$
hypersurface $\Sigma_t$, $\Lapse$ the lapse-function and $\beta^i$ the
shift-vector.
The extrinsic curvature is defined as
\begin{equation}\label{eq:K-defn}
\KW_{\ell m}\equiv \frac{1}{2} {\cal L}_{^{(4)}n}\; g_{\ell m},
\end{equation}
where ${\cal L}_{^{(4)}n}$ is the Lie derivative along the
hypersurface unit-normal $^{(4)}n^\mu$. In Eq.~(\ref{eq:K-defn}), as
throughout this paper, we employ the sign convention of
Wald~\cite{Wald}, resulting in a {\em positive} mean curvature $\TrKW
\equiv g_{ij}\KW^{ij}$ for the cases considered. This sign-convention
differs from the conventions of Misner, Thorne \& Wheeler~\cite{MTW},
which are more commonly used in numerical relativity (as, for
instance, in Ref.~\cite{Cook2004}).

Einstein's vacuum constraint equations are
\begin{equation}\label{eq:EinsteinHamiltonian}
\mathcal{R}+\TrKW^2-\KW_{ij}\;\KW^{ij}=0,
\end{equation}
and
\begin{equation}
\label{eq:EinsteinMomentum}
\nabla_j (\KW^{ij}-g^{ij}\TrKW)=0,
\end{equation}
where ${\nabla}_i$ is the covariant derivative with respect to
${g}_{ij}$, and $\mathcal{R}$ is the Ricci scalar associated with
$g_{ij}$.

We now perform a conformal transformation which plays a dual role. On
one hand, it allows a conformal compactification
(Penrose~\cite{Penrose1965}), placing $\ScriPlus$ at a {\em finite}
value of a compactified radial coordinate, absorbing the resulting
metric singularities into the conformal factor. On the other hand, it
allows the Einstein constraints to be recast as elliptic equations
following the standard procedure of the conformal
method~\cite{Murchadha-York:1974b,Pfeiffer2003b}.

The conformal
  metric $\tilde g_{ij}$ and conformal factor $\Omega$ are given by 
\begin{equation}\label{eq:ConfMetric}
g_{ij} = \Omega^{-2}~\tilde{g}_{ij},~~~~~g^{ij} =
\Omega^{2}~\tilde{g}^{ij}.
\end{equation}
The conformal metric is assumed to be regular at $\ScriPlus$ in
compactified coordinates. This implies that $\Omega=0$ at
$\ScriPlus$ since, in compactified coordinates, the physical
metric is singular there.  Comparing Eq.~(\ref{eq:ConfMetric}) with
the more widely used definition $g_{ij}=\psi^4\tilde g_{ij}$, we see
that $\Omega=\psi^{-2}$.  The advantage of the
definition~(\ref{eq:ConfMetric}) is that $\Omega$ is {\em finite} at
$\ScriPlus$.  The equations in the rest of this section mirror the
standard conformal method~\cite{Murchadha-York:1974b,Pfeiffer2003b}, 
with the replacement $\Omega\leftrightarrow\psi^{-2}$.

The trace-free extrinsic curvature is defined as 
\begin{equation}
\label{eq:Aij}
\AW^{ij}=\KW^{ij}-\frac{1}{3}\;g^{ij}\TrKW,
\end{equation}
where $\KW^{ij}=g^{i \ell} g^{jm}\KW_{\ell m}$.  The physical
trace-free extrinsic curvature $\AW^{ij}$ and its conformal
counterpart $\tilde{\AW}^{ij}$ are related by
\begin{equation}\label{eq:CAij}
\AW_{ij} = \Omega ~\tilde{\AW}_{ij},~~~~~\AW^{ij} = \Omega^5
~\tilde{\AW}^{ij}.
\end{equation}

Substituting Eq.~(\ref{eq:Aij}) into Eq.~(\ref{eq:EinsteinMomentum})
and using the CMC condition $\nabla_j \TrKW =
0$ gives
\begin{equation}
\nabla_j \AW^{ij}=0.
\end{equation}
This can be rewritten as
\begin{equation}\label{eq:CMom}
\Omega^5 \tilde{\nabla}_j \tilde{\AW}^{ij} = 0,
\end{equation}
where $\tilde{\nabla}_i$ is the covariant derivative with respect to
$\tilde{g}_{ij}$. Because of the CMC condition, the momentum
constraint has decoupled from the Hamiltonian constraint, and can be
solved first. Standard methods for solving the momentum constraint are
given in Ref.~\cite{York1973}. In this paper, we assume a flat
conformal metric $\tilde{g}_{ij}$.

Substituting the solution $\tilde{\AW}^{ij}$ of the momentum
constraint into the Hamiltonian
constraint~(\ref{eq:EinsteinHamiltonian}) and expressing it in terms
of conformal quantities, one finds 
\begin{equation}
  \label{eq:HamiltonianConstraint} \tilde{\nabla}^2 \Omega\, -\,
  \frac{3}{2 \Omega}\big(\tilde{\nabla}\Omega\big)^2 +
  \frac{\Omega}{4}\,\tilde{\mathcal{R}}+ \frac{\TrKW^2}{6\,\Omega} -
  \frac{\Omega^5}{4}\,\tilde{\AW}_{ij}\tilde{\AW}^{ij}=0.
\end{equation} 
Here $\tilde{\nabla}^2$ denotes the covariant Laplacian
with respect to $\tilde{g}_{ij}$ and $\tilde{\mathcal{R}}$ is the
scalar curvature of $\tilde{g}_{ij}$. Note that some terms in
Eq.~(\ref{eq:HamiltonianConstraint}) are singular at $\ScriPlus$
where $\Omega=0$. Assuming $\tilde{\AW}_{ij}$ is finite at
$\ScriPlus$, any regular solution satisfying the boundary condition
\begin{equation}\label{eq:OmegaOnScriPlusI}
  \left.\Omega\right|_{\ScriPlus} = 0 \end{equation} must also satisfy
\begin{equation}\label{eq:OmegaOnScriPlusII}
  \left.\left(\tilde{\nabla}\Omega\right)^2\right|_{\ScriPlus} =
  \left(\frac{\TrKW}{3}\right)^2. \end{equation}

\subsection{The Schwarzschild black hole in CMC slicing}
\label{Sec:Schwarzschild}
Let us now discuss this initial value formalism in the particular case
of a Schwarzschild black hole, recasting already established
results~\cite{Brill1980,Malec2003,Malec2009} in our language. In
spherical coordinates $(r,\theta,\phi,t)$, with $t$ constant on CMC
slices and $r$ the areal radius, the metric is given
by~\cite{Brill1980,Malec2003,Malec2009}
\begin{align}
\label{eq:SchwarzCMC}
ds^2=& -\left(1-\frac{2M}{r}\right)dt^2+\frac{1}{f^2}dr^2
-\frac{2 a}{f}dt\,dr\\
&+r^2\left(d\theta^2+\sin^2\theta\,d\phi^2\right),\nonumber
\end{align}
where the functions $f=f(r)$ and $a=a(r)$ are
\begin{equation}\label{eq:Def-f-a}
f(r)=\left(1-\frac{2M}{r}+a^2\right)^{1/2},
\quad\;\; a(r)=\frac{\TrKW r}{3}-\frac{C}{r^2}.
\end{equation}
With these definitions, $a(r)={}^{(4)}n^r$, where $^{(4)}n^r$ is the
radial component of the hypersurface-normal $^{(4)}n^\mu$. The
constant $M$ is the mass of the black hole. The constant $C$
represents an additional one-parameter degree of freedom in the choice
of spherically symmetric CMC hypersurfaces in the Schwarzschild
metric. The lapse-function $\Lapse$ of the
metric~(\ref{eq:SchwarzCMC}) equals $f$, and the only non-vanishing
component of the shift is $\beta^r=-af$.

The coordinate transformation between Schwarzschild coordinates
$(r,\theta,\phi,T)$ and the CMC coordinates $(r,\theta,\phi,t)$
(again, see~\cite{Brill1980,Malec2003,Malec2009}) is
\begin{eqnarray}
t&=&T-\int{\frac{a(r)}{\left(1-\frac{2M}{r}\right)f(r)}dr}\\\nonumber
&=&u+\int{\frac{dr}{\left(1-\frac{2M}{r}\right)}} -
\int{\frac{a(r)}{\left(1-\frac{2M}{r}\right)f(r)}dr},
\end{eqnarray}
where $u$ is the Eddington-Finkelstein retarded null coordinate. With
$\TrKW>0$, constants of integration can be chosen so that
$t\rightarrow u$ as $r \rightarrow \infty$.

The hypersurfaces $\Sigma_t$ of constant coordinate value $t$ will
play a central role in this paper. For $\TrKW=0$, these hypersurfaces
extend to space-like infinity. For $\TrKW\neq 0$, $\Sigma_t$ becomes
asymptotically null for large radius. If $\TrKW>0$, this hypersurface
intersects future null-infinity, because $1/f \rightarrow 0$ and $a/f
\rightarrow +1$ in the limit $r\to\infty$. For $\TrKW<0$, it
intersects past null infinity. We are interested in hypersurfaces
approaching future null-infinity, and will therefore require
\begin{equation}\label{eq:K-sign}
\TrKW>0
\end{equation}
throughout this paper.

The constant $C$ determines whether the $\Sigma_t$ hypersurface
intersects the black hole horizon or the white hole horizon.
Considering radial null rays on the horizon, $r=2M$, we find that for
\begin{equation}\label{eq:C-inequality}
C>\frac{8}{3}\TrKW M^3,
\end{equation}
$\Sigma_t$ intersects the black hole horizon (i.e. $^{(4)}n^r < 0$ for
$r \le 2M$). If the inequality is reversed, the hypersurface enters
the white hole region, where the light cone is tilted toward
increasing $r$, and excision is is not possible without allowing
causal propagation from the excision boundary to the interior of the
computational domain.

As with any spherically symmetric metric, a radial coordinate
transformation $r\!\to\! R=R(r)$ can be used to make the spatial
sector of Eq.~(\ref{eq:SchwarzCMC}) conformally flat, i.e.
$ \Omega^{-2}\left[dR^2 + R^2(d\theta^2+\sin^2\theta d\phi^2)\right]$.
Here,
\begin{equation}
\Omega=\frac{R}{r},
\end{equation}
and $R$ is determined by the ordinary differential equation
\begin{equation}\label{eq:rR-mapping}
\frac{dR}{dr}=\frac{R}{rf}.  
\end{equation}
Because $f\sim r$ for large $r$, $R$ remains {\em finite} as
$r\to\infty$.  Denoting its limiting value by $R_+$, we find
\begin{equation}
\label{eq:ROverRmax}
\frac{R(r)}{R_+} = 
\exp\left(- \int_{r}^\infty \frac{dr'}{r' f(r')}\right).
\end{equation}
As $r\to\infty$ (or equivalently $R\to R_+$), $\Omega\to 0$, in
agreement with the boundary condition Eq.~(\ref{eq:OmegaOnScriPlusI}).

Finally, transforming $(R,\theta,\phi)$ to Cartesian coordinates
$x^i$, we can express the space-time metric
Eq.~(\ref{eq:SchwarzCMC}) as
\begin{equation}
\label{eq:Schwarz2}
ds^2 = \Omega^{-2} \left[ -\tilde{\alpha}^2
dt^2+\delta_{ij}(dx^i\!+\!\tilde{\beta}^i dt)(dx^j\!+\!\tilde{\beta}^j dt)\right].
\end{equation}
Here the conformal lapse $\tilde\Lapse$ and conformal shift
$\tilde\beta^i$ are given by
\begin{equation}
\tilde{\alpha}=\Omega f~~~
\text{and}~~~\tilde{\beta}^i=-\Omega a\,n^i,
\end{equation}
where $n^i=x^i/R$. Because $\Omega\sim R_+/r$ whereas $f\sim a\sim r$
as $r\to\infty$, $\tilde\Lapse$ and $\tilde\beta^i$ are {\em finite}
at $\ScriPlus$. Therefore, the conformal space-time metric inside the
square brackets in Eq.~(\ref{eq:Schwarz2}) is regular, in addition to
its spatial part being flat.

The trace-free extrinsic curvature of the $\Sigma_t$ hypersurface in
the coordinates $(t, x^i)$ takes the form $\AW^{ij}=\Omega^5
\tilde{\AW}^{ij}$, with
\begin{equation}\label{eq:Atilde-CMC}
\tilde{\AW}^{ij}=\frac{C}{R^3} \Big( 3 n^i n^j - \delta^{ij} \Big).
\end{equation}
The sign difference in Eq.~(\ref{eq:Atilde-CMC}) relative to earlier
work (e.g. Eq.~(52d) of~\cite{Cook2004}) arises because of the
different sign-convention for $\AW_{ij}$ [see the discussion after
  Eq.~(\ref{eq:K-defn})].  The conformal trace-free extrinsic
curvature satisfies the momentum constraint Eq.~(\ref{eq:CMom}).
Indices on conformal spatial tensors such as $\tilde{\AW}^{ij}$ are
raised or lowered with the conformal spatial metric; in our
coordinates, this is simply the Kronecker delta, so the components of
such tensors are identical irrespective of the index-location.

We finish this section by noting that Eq.~(\ref{eq:ROverRmax})
determines only $R(r)/R_+$; the freedom remains to rescale $R$ 
 by a real constant $\Rrescale$,
\begin{subequations}\label{eq:Rrescaling}
\begin{equation}
R\to \Rrescale R.
\end{equation}
This rescaling induces further rescalings,
\begin{align}
\Omega&\to \Rrescale \Omega,\\
\tilde{\AW}^{ij}&\to\Rrescale^{-3}\tilde{\AW}^{ij},\\
\label{eq:RrescaleA_ij}
\tilde{\AW}_{ij}&\to\Rrescale^{-3}\tilde{\AW}_{ij}.
\end{align}
\end{subequations} 
Eqs.~(\ref{eq:Rrescaling}) represent a coordinate transformation and
do not affect the physical initial data.  For the {\em zero} mean
  curvature case, $\TrKW=0$, the hypersurface asymptotes to spatial
  infinity, where it is natural to impose the boundary condition
  $\Omega \rightarrow 1$, so $R/r \rightarrow 1$ as $R \rightarrow
  \infty$.  With $\TrKW>0$, the coordinate scale is set by the
  arbitrary choice of the value of $R$ at future null infinity, $R_+$.
  Only after the Hamiltonian constraint is solved can $R$ and $\Omega$
  be rescaled to make the maximum value of $\Omega$ equal one. If
  $\TrKW$ is very close to zero, this
  rescaling makes $\Omega$ very close to the $\TrKW=0$ solution almost
  everywhere.

%%%%%%%%%%%%%%%%%%%%%%%%%%%%%%%%%%%%%%%%%%%%%%%%%%%%%%%%%%%%%%%%
\subsection{Minimal Surfaces \& Trumpets}
\label{sec:MinimalSurfacesAndTrumpets}
%%%%%%%%%%%%%%%%%%%%%%%%%%%%%%%%%%%%%%%%%%%%%%%%%%%%%%%%%%%%%%%%

Later in this paper, we will construct black hole initial data with
minimal surface boundary conditions in the interior of the black
hole(s). As preparation, let us discuss the presence and location of
minimal surfaces in the hypersurfaces $\Sigma_t$ of the metric
Eq.~(\ref{eq:SchwarzCMC}). Our treatment here extends earlier similar
discussions~\cite{Brill1980,Malec2003,Malec2009}.

The sphere $r=$constant is a minimal 2-sphere within $\Sigma_t$ if the
function $f$ defined in Eq.~(\ref{eq:Def-f-a}) vanishes. This can be
seen most easily by noting that $r$ is the areal radius, and that from
Eq.~(\ref{eq:rR-mapping}), $dr/dR=f\,r/R$, which vanishes for $f=0$.
For any values $M>0$, $K$ and $C$, $f(r)>0$ for $r>2M$. Furthermore,
the inequalities Eqs.~(\ref{eq:K-sign}) and (\ref{eq:C-inequality})
imply that $f$ is strictly positive, $f>0$, at the black hole horizon
$r=r_H\equiv2M$. Therefore, a minimal surface in $\Sigma_t$ will
always lie inside the horizon, if it exists.

Vacuum general relativity possesses a rescaling freedom, and this
freedom will be inherited by the function $f$ and any equations that
describe minimal surfaces.  The most common way to incorporate this
freedom is to rescale all dimensionful quantities by the mass $M$.  In
particular, Eq.~(\ref{eq:Def-f-a}) can be written in terms of the
dimensionless variables $r/M$, $\TrKW M$ and $C/M^2$ as
\begin{equation}
f^2=1-\frac{2}{r/M}+\left[\frac{(\TrKW
    M)(r/M)}{3}-\frac{C/M^2}{(r/M)^2}\right]^2.
\end{equation}
Using $M$ to make dimensionless variables is fine for Schwarzschild, 
but in the absence of spherical symmetry, the mass $M$ is not known in advance.  
On CMC hypersurfaces, $\TrKW$ is a free parameter and can be used to form an 
alternative and more widely applicable set of dimensionless variables, 
$\TrKW r$, $\TrKW^2\sqrt{C}$, and $\TrKW M$, such that
\begin{equation}\label{eq:f-Jim-Params}
  f^2=1-\frac{2 \TrKW M}{\TrKW r} + \left[ \frac{\TrKW
    r}{3}-\frac{\TrKW^2 C}{(\TrKW r)^2}\right]^2.
\end{equation}
A minimal surface at radius $r_{\rm ms}$ satisfies
\begin{equation}\label{eq:f=zero}
1-\frac{2 \TrKW M}{\TrKW r_{\rm ms}} + \left[ \frac{\TrKW r_{\rm
      ms}}{3}-\frac{\TrKW^2 C}{(\TrKW r_{\rm ms})^2}\right]^2=0.
\end{equation}
This equation will play a central role in the remainder of this
section.

\begin{figure}
\includegraphics[width=0.9\columnwidth]{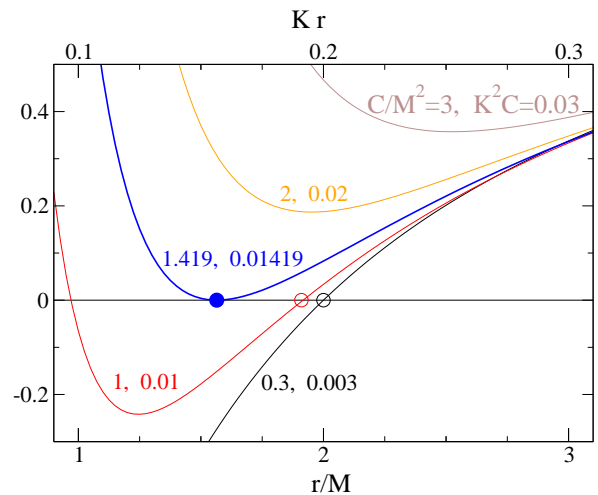}
\caption{\label{fig:RadialFunctionFSquared} Radial function $f(r)$ for
  different values of $\TrKW$ and $C$. Plotted is the square $f^2$ for
  $\TrKW M=1/10$ and several different choices for $C$.  Each curve is
  labeled by its value of $C/M^2$ and its value for $\TrKW^2C$. The
  bottom axis shows radius in units of $r/M$, and the top axis in
  units of $\TrKW r$. }
\end{figure}

Fig.~\ref{fig:RadialFunctionFSquared} plots $f^2$ for different values
of $C/M^2$ for a fixed value of $\TrKW M$. For large $C/M^2$ (or
equivalently $\TrKW^2C$), no root and no minimal surface exists. At
some critical value $C_{\rm T}/\MT^2$, $f$ touches zero, indicated by
the filled circle. For $C/M^2<\CT/\MT^2$, minimal surfaces exist at
radii $r_{\rm ms}/M$, as indicated by the open circles in
Fig.~\ref{fig:RadialFunctionFSquared}. The critical value $\CT/\MT^2$
delineates the region of parameters for which $\Sigma_t$ contains a
minimal surface. At this critical point, $f$ varies {\em linearly} in
$r-\rT$, passing through zero at $\rT$. From
Eq.~(\ref{eq:SchwarzCMC}), the radial proper separation within the
slice is $ds=dr/f$. Because $f$ vanishes linearly at $\rT$, this point
is an infinite proper distance away from any point $r>\rT$: this
configuration is often called a {\em
  trumpet}~\cite{Hannam2006,Hannam2007d,Hannam2008,Hannam2009}. In
contrast, away from the critical point (i.e. at the open circles in
Fig.~\ref{fig:RadialFunctionFSquared}), $f$ approaches zero
proportionally to $\sqrt{r-r_{\rm ms}}$, and the proper distance to
the minimal surface is finite.

For trumpets, $f^2=0$ and $\partial_r f^2=0$; the parameter values
that lead to trumpets can be written down in terms of the areal radius
of the trumpet, $r_{\rm T}$:
\begin{subequations}
\label{eq:KCTrumpet}
\begin{align}\label{eq:KTMT_vs_rT}
\KT \MT &=
\frac{2\rT/\MT-3}{(\rT/\MT)\sqrt{(\rT/\MT)(2-\rT/\MT)}},\\ 
\frac{\CT}{\MT^2} &= 
\frac{(\rT/\MT)^2(3-\rT/\MT)}{3\sqrt{(\rT/\MT)(2-\rT/\MT)}},
%
% xmgrace EXPRESSIONS
%  (3-2*$t)/($t*sqrt($t*(2-$t)))
%  ($t)^2*(3-$t)/(3*sqrt($t*(2-$t)))
%
\end{align}
\end{subequations}
where $3/2<\rT/\MT<2$. 

\begin{figure}
\includegraphics[scale=0.5]{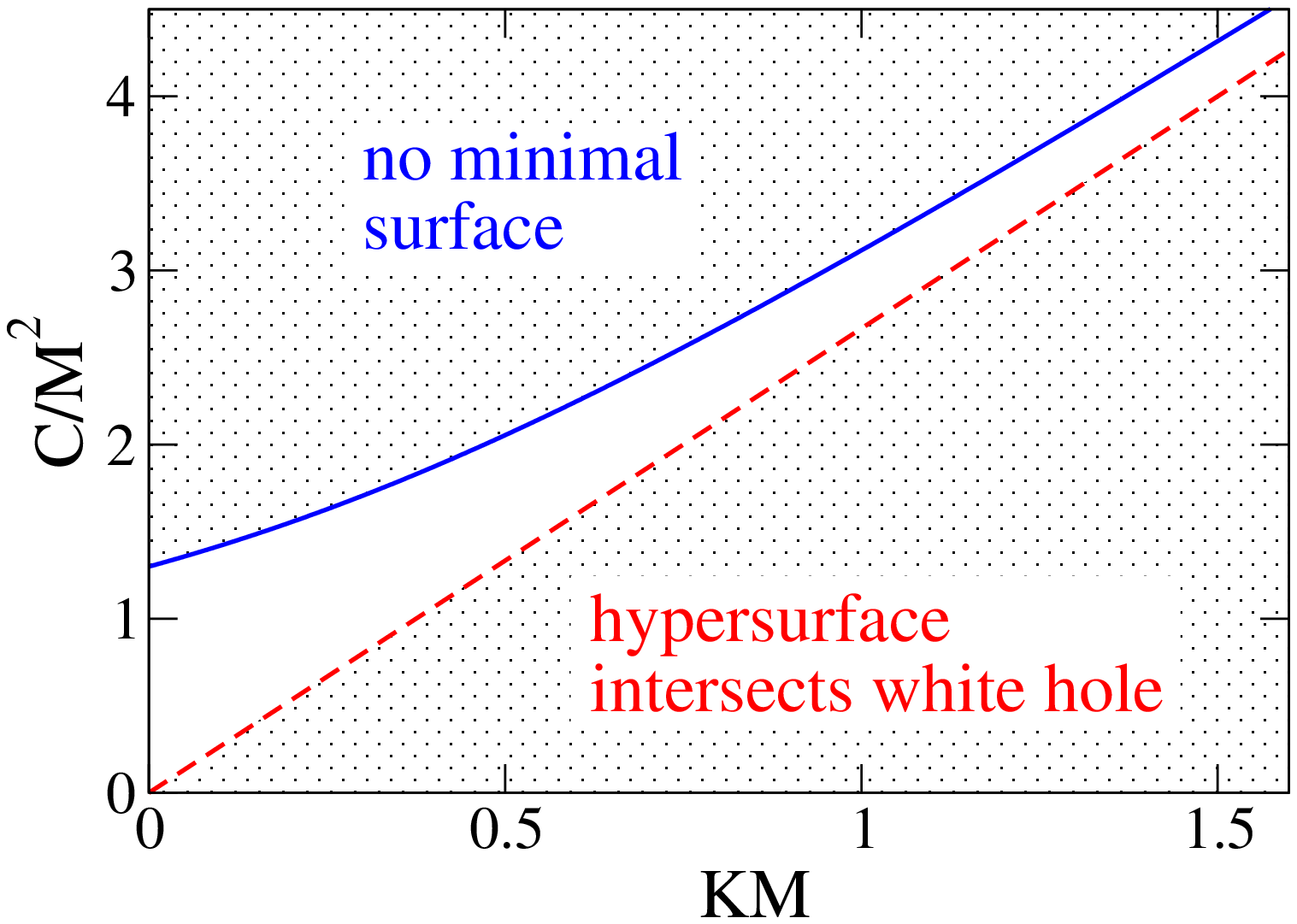}\\[2em]
\includegraphics[scale=0.5]{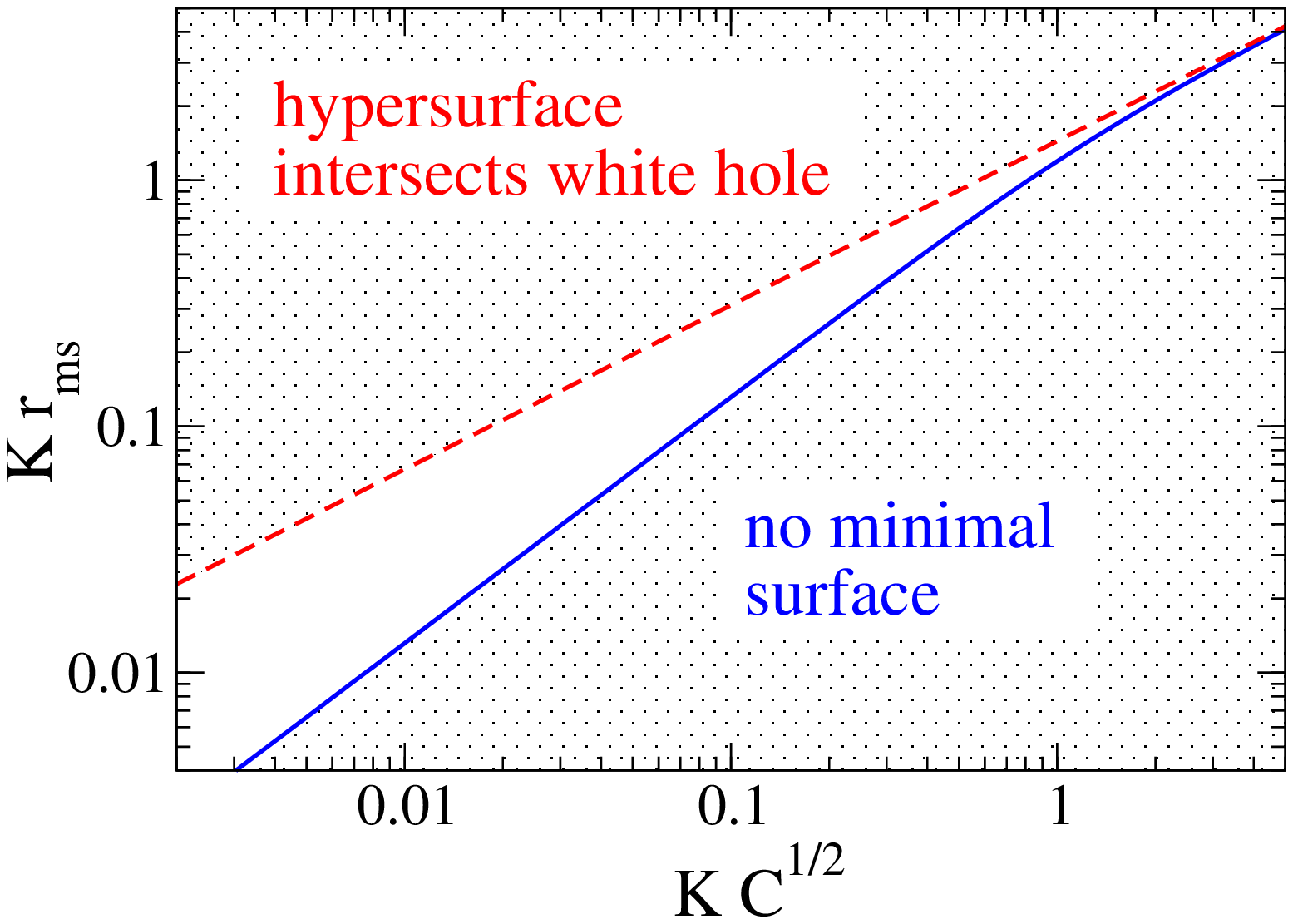}
\caption{\label{fig:K-C-Plane}Parameter choices that result in
  hypersurfaces $\Sigma_t$ containing a minimal surface.  The two
  panels correspond to two different choices of dimensionless
  variables.  Parameter values on the blue solid lines represent trumpet
  configurations. }
\end{figure}

The preceding discussions determine the region of parameters $M,
\TrKW, C$, for which minimal surfaces exist. Taking the scaling
invariance into account, this region can be represented on a
two-dimensional plot as given in Fig.~\ref{fig:K-C-Plane}. The blue
solid and red dashed lines in the top panel, for instance, are given
by a parametric plot of Eqs.~(\ref{eq:KCTrumpet}) and
Eq.~(\ref{eq:C-inequality}). The unshaded wedge-shaped region between
these two lines represents the allowed parameter choices which lead to
a CMC hypersurface containing a minimal surface and intersecting the
black hole.

\begin{figure}
\includegraphics[width=0.88\linewidth]{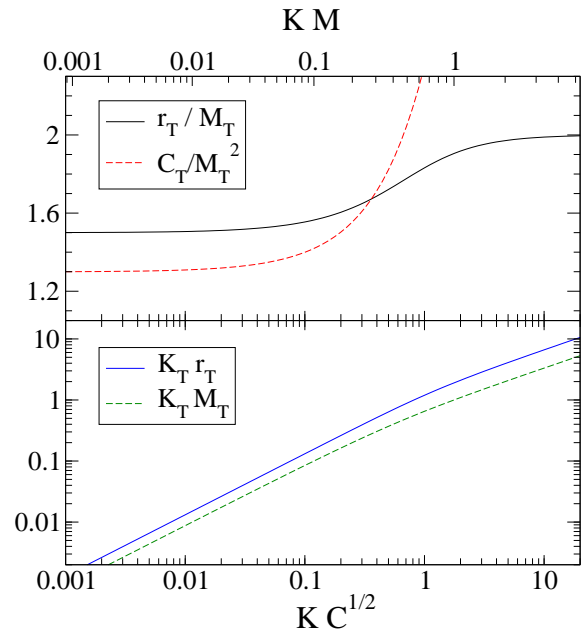}
\caption{\label{fig:trumpets}Properties of the trumpet hypersurfaces,
  parametrized by the dimensionless parameter $\TrKW C^{1/2}$.}
\end{figure}

Alternatively, one can compute trumpet-configurations using the
dimensionless variables indicated in Eq.~(\ref{eq:f-Jim-Params}).
From the equations $f^2=0$ and $\partial_r f^2=0$, one eliminates
$\TrKW M$ to obtain a third order polynomial that relates $\KT\rT$ and
$\KT\CT^{1/2}$.  This polynomial has only one positive real root, which
for $\KT^2\CT<2/3$ can be written as
\begin{equation}\label{eq:KrTrumpet}
  (\KT\rT)^2 = \frac{\frac{3}{2}\KT^2\CT}
  {\cos\left[\frac{1}{3}\arccos\left(\frac{3}{2}\KT^2\CT\right)\right]}.
\end{equation}
For $\KT^2\CT>2/3$, the trigonometric functions in
Eq.~(\ref{eq:KrTrumpet}) should be replaced by their hyperbolic
counterparts. To find the trumpet solution, substitute $\KT\rT$ back
into Eq.~(\ref{eq:f-Jim-Params}):
\begin{equation}\label{eq:KMTrumpet}
  \KT\MT=
  \frac{\KT\rT}{2}\left(1+\left[\frac{\KT\rT}{3}-\frac{\KT^2\CT}{(\KT\rT)^2}\right]^2\right).
\end{equation}
Substituting $\KT\rT$ from Eq.~(\ref{eq:KrTrumpet}), $\KT\MT$ is a
function of $\KT^2\CT$ alone. Finally, dividing
Eq.~(\ref{eq:KrTrumpet}) by Eq.~(\ref{eq:KMTrumpet}) yields the value
of $\rT/\MT$. All these parameters for trumpet hypersurfaces are
plotted in Fig.~\ref{fig:trumpets}. To make easy contact with
dimensionless quantities normalized by $M$, the top horizontal axis of
this plot is labeled by $\TrKW M$. For $\TrKW C^{1/2}\ll 1$, the data
plotted in the lower panel of Fig.~\ref{fig:trumpets} is proportional
to $\KT\CT^{1/2}$:
\begin{align}
  \KT\rT&=3^{1/4} \KT\CT^{1/2}+{\cal O}\left(
    (\KT^2\CT)^{3/2}\right),\\ \KT\MT&=\frac{2}{3^{3/4}}
  \KT\CT^{1/2}+{\cal O}\left( \KT^2\CT\right).
\end{align}

For given choices $\TrKW$ and $\TrKW^2C$, minimal surfaces only exist
at radii $r_{\rm ms}$ larger than $\rT$ given by
Eq.~(\ref{eq:KrTrumpet}). This is indicated by the solid blue line in
the lower panel of Fig.~\ref{fig:K-C-Plane}.

To obtain an upper limit on $\TrKW r_{\rm ms}$, we recall that all minimal
surfaces must lie inside the horizon, $r_{\rm ms}<r_H=2M$.  Combining this
with Eq.~(\ref{eq:C-inequality}) results in
\begin{equation}\label{eq:maxKrms}
  K r_{\rm ms}< 2MK < \left(3 K^2 C\right)^{1/3},
\end{equation}
which is indicated by the red dashed line in the lower panel of
Fig.~\ref{fig:K-C-Plane}.

\begin{figure}
 \includegraphics[width=0.99\linewidth]{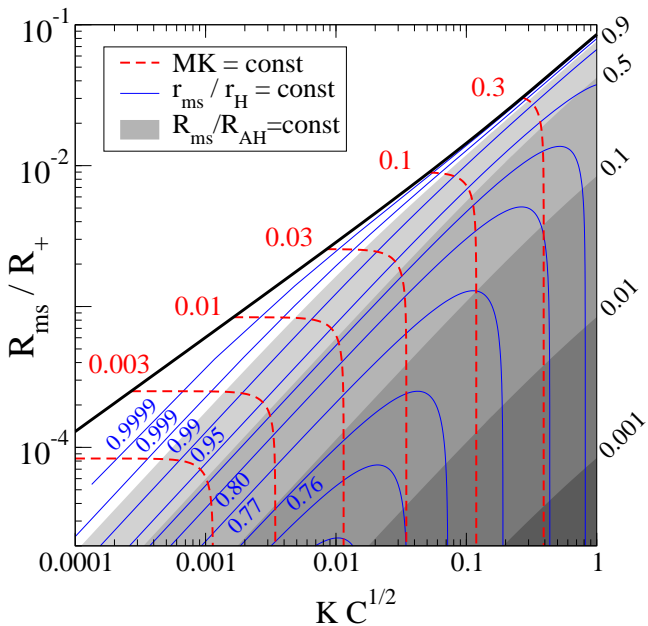}
 \caption{\label{fig:ContoursKM}Properties of CMC hypersurfaces viewed
   in the $R_{\rm ms}/R_+$ vs. $\TrKW C^{1/2}$ plane. CMC
   hypersurfaces with a minimal surface must lie below the thick black
   line, and as this line is approached, the minimal surface approaches
   the black hole horizon. The dashed red lines are lines of constant
   $MK$, with values given by the red numbers next to the lines. The
   thin blue lines represent constant values of $\gamma=r_{\rm
     ms}/r_H$. The shaded areas are contours of constant value of
   $R_{\rm ms}/R_{\rm AH}$; the shade of grey changes, from top to
   bottom, at values 0.9, 0.5, 0.1, 0.01, and 0.001. Trumpet
   hypersurfaces represent the limit $R_{\rm ms}/R_+\to 0$.}
\end{figure}

Finally, Fig.~\ref{fig:ContoursKM} presents another view of the
two-dimensional set of ``good'' parameter choices that we first
indicated in Fig.~\ref{fig:K-C-Plane}.  As in the lower panel of
Fig.~\ref{fig:K-C-Plane}, we shall use $\TrKW C^{1/2}$ to parametrize
the horizontal axis.  However, we use Eq.~(\ref{eq:ROverRmax}) in the
form
\begin{equation}\label{eq:RmsOverRplus-u}
\ln\frac{R_{\rm ms}}{R_+}=-\!\int_0^ {
%(Kr_{\rm ms})^{-1}
1/(\TrKW r_{\rm ms})
%\frac{1}{Kr_{\rm ms}}
}\!\!\!\!\!\!\frac{du}{\sqrt{u^2-2\TrKW M u^3+\left(\frac{1}{3}-\TrKW^2 C u^3\right)^2}}
\end{equation}
to convert the vertical axis to the ratio $R_{\rm ms}/R_+$ of
conformal radius of the minimal surface and the conformal radius of
$\ScriPlus$.  Fig.~\ref{fig:ContoursKM} shows the following data:
First, the {\bf thick black line} corresponds to parameters for which
the minimal surface coincides with the black hole horizon.  Setting
$\TrKW r_{\rm ms}=2\TrKW M$ in Eq.~(\ref{eq:f=zero}) and solving for
$\TrKW r_{\rm ms}$ shows that this line is parametrized by
\begin{equation}
\TrKW r_{\rm ms} = (3\TrKW^2 C)^{1/3},
\end{equation}
with $\TrKW r_{\rm ms}$ mapped to $R_{\rm ms}/R_+$ by
Eq.~(\ref{eq:RmsOverRplus-u}).  The {\bf red dashed lines} are
contours of constant values $\TrKW M$.  Each of these lines is
obtained by keeping $\TrKW M$ fixed and varying $\TrKW r_{\rm ms}$
between its lower bound, the trumpet value $\KT\rT$ (obtained from
inverting Eq.~(\ref{eq:KTMT_vs_rT})) and its maximal value $2\TrKW
M$.  For each choice of $\TrKW M$ and $\TrKW r$, Eq.~(\ref{eq:f=zero})
is solved for $\TrKW^2C$, and the resulting data plotted as a
parametric plot.  Finally, the {\bf thin blue lines} in
Fig.~\ref{fig:ContoursKM} represent lines of constant ratio $\TrKW
r_{\rm ms}/(2\TrKW M)=r_{\rm ms}/r_H\equiv\gamma$, i.e. lines where the areal radius
of the minimal surface is a fixed fraction of the areal radius of the
horizon.  Replacing $\TrKW r_{\rm ms}$ by $2\gamma \TrKW M$ in
Eq.~(\ref{eq:f=zero}), we can solve for $\TrKW C^{1/2}$ as a function
of $\TrKW M$.  The thin blue lines are then obtained as a parametric plot
$(\TrKW C^{1/2}, \TrKW r)$ as $\TrKW M$ is varied.

Trumpet initial conditions are obtained from Fig.~\ref{fig:ContoursKM}
through the limit $R_{\rm ms}/R_+\to 0$, i.e. by going ``down''. Note
that the red $\TrKW M=$constant contours become vertical in this
limit. Their value as a function of $\TrKW C^{1/2}$ is then given in
the lower panel of Fig.~\ref{fig:trumpets}.

The significance of the axes employed in Fig.~\ref{fig:ContoursKM} is
that both axes represent quantities that are freely specifiable when
constructing CMC hypersurfaces within the initial value formalism of
general relativity. One use of Fig.~\ref{fig:ContoursKM} is to first
pick values of $\TrKW$ and $M$, fixing a particular contour of $\TrKW
M$. One then chooses how trumpet-like the initial conditions should
be; that is, how far to go down along the contour. Alternatively,
one can choose a certain ratio $\gamma=r_{\rm ms}/r_H$. In either
case, one can then read off the corresponding values of $R_{\rm
  ms}/R_+$ and $\TrKW \sqrt{C}$ to get the remaining initial value
parameters.

\subsection{CMC initial data for single black holes}
\label{Sec:SchwarzIVP}
The properties of CMC slices of the Schwarzschild spacetime, as
described in Secs.~\ref{Sec:Schwarzschild}
and~\ref{sec:MinimalSurfacesAndTrumpets}, mesh nicely with the
conformal method of solving the Einstein constraint equations, which
was outlined in Sec.~\ref{Sec:IVP}. We shall first discuss the
spherically symmetric case, for which there is a one-to-one
correspondence. Subsequently, we will generalize to a single spinning
or boosted black hole.

\subsubsection{Spherical symmetry}
\label{SubSec:SpherSym}
The CMC metric is conformally flat, so we shall use a flat conformal
metric for the initial value problem,
\begin{equation}\label{eq:ConfFlat}
\tilde g_{ij}=f_{ij},
\end{equation}
where $f_{ij}$ is the flat space metric. The extrinsic curvature of
the CMC metric has the correct scaling with conformal factor [compare
Eqs.~(\ref{eq:CAij}) and (\ref{eq:Atilde-CMC})], so we shall adopt
Eq.~(\ref{eq:Atilde-CMC}) as the freely specifiable tracefree
extrinsic curvature, with the constant $C$ yet to be determined.  The
radial coordinate $R$ ranges from a finite value $R_+$ representing
$\ScriPlus$ to some smaller value, for instance $R_{\rm ms}$ at a
minimal surface (assuming a minimal surface exists), so we shall adopt
a computational domain with inner radius $R_1$ and outer radius $R_2$.
At $\ScriPlus$, the conformal factor vanishes, resulting in the
boundary condition
\begin{equation}
\Omega=0,\qquad R=R_2,
\end{equation}
which identifies $R_2$ with $R_+$.  At the inner boundary, we shall
impose a minimal surface boundary condition,
\begin{equation}\label{eq:MinimalSurface-BC-1}
\frac{d\Omega}{dR}=\frac{\Omega}{R},\qquad R=R_1,
\end{equation}
so that $R_1$ will coincide with $R_{\rm ms}$.  

With the choices
Eq.~(\ref{eq:ConfFlat})--(\ref{eq:MinimalSurface-BC-1}), we are now
left with choosing the four numbers $\{\TrKW, C, R_1, R_2\}$.
Solution of the Hamiltonian constraint
Eq.~(\ref{eq:HamiltonianConstraint}) will then result in a complete
initial data set with a certain mass $M$.  Fig.~\ref{fig:ContoursKM}
is useful for informed choices for the numbers $\{\TrKW, C, R_1,
R_2\}$. For instance, we can first decide on a mass $M$ (say, $M=1$)
and a mean curvature $\TrKW$ (say, $\TrKW=0.01$).  This selects one of
the red-dashed curves in Fig.~\ref{fig:ContoursKM}.  We can now choose
a suitable value of $r_{\rm ms}/r_H$ by considering the intersection
of the red dashed lines with the blue contours (say, $r_{\rm
  ms}/r_H=0.9$), and read off the values for $\TrKW C^{1/2}$ and
$R_{\rm ms}/R_+$ (in our example $\TrKW C^{1/2}=0.0105$, $R_{\rm
  ms}/R_+=0.000554$), which determine the values for $C$ and
$R_1/R_2$.  An overall scaling of $R_1$ and $R_2$ remains, because the
coordinate transformation $r\to R$ for CMC slices is determined only
up to an overall rescaling (see the discussion after
Eq.~(\ref{eq:ROverRmax})).  Thus, we are free to set, for instance,
$R_1=1$.

\subsubsection{Single black hole with spin \& boost}
\label{SubSec:BhPlusSpinBoost}

A relatively simple class of non-spherically symmetric initial data on
maximal slices, with $\TrKW=0$, was proposed by Bowen and
York~\cite{bowen79,Bowen-York:1980}. It assumes conformal flatness of
the spatial metric and a solution $\tilde A_{ij}$ of the conformal
momentum constraint Eq.~(\ref{eq:CMom}) characterized by a ``spin''
vector $S^i$ and two ``boost'' vectors $P^i$, $Q^i$. On the
asymptotically flat maximal slices, with the boundary condition
$\Omega \rightarrow 1$ at spatial infinity, $S^i$ is in fact the
physical angular momentum and $P^i$ is the physical linear momentum of
the system, as defined by Arnowitt-Deser-Misner (ADM) surface
integrals at spatial infinity~\cite{ADM}. The second boost vector
$Q^i$ is introduced to allow for inversion symmetry about a minimal
surface, and can be thought of as the three-momentum of the black hole
as viewed from the asymptotically flat space on the other side of the
Einstein-Rosen bridge associated with the minimal surface.

Since Eq.~(\ref{eq:CMom}) is linear and identical on maximal and CMC
slices, we can add the Bowen-York terms to Eq.~(\ref{eq:Atilde-CMC}),
with the result\footnote{The sign differences between
  Eq.~(\ref{eq:GeneralMomentumTensor}) and earlier papers arise
  because of the different sign convention for $\AW_{ij}$ (see
  Eq.~(\ref{eq:K-defn})).}:
\begin{eqnarray}
  \label{eq:GeneralMomentumTensor}
  \tilde{\AW}_{ij}
  &=&\frac{C}{R^3} \left[ 3\,n_in_j-\delta_{ij}\right]\\
  &&-\frac{3}{R^3} \left[\varepsilon_{ik\ell}\,S^kn^\ell n_j
    +\varepsilon_{jk\ell}\,S^kn^\ell n_i \right]\nonumber\\ 
  && -\frac{3}{2R^2} \left[P_i n_j+P_j n_i+
    P^kn_k\,(n_in_j-\delta_{ij})\right] \nonumber\\ 
  &&+\frac{3}{2R^4} \left[ Q_in_j+Q_jn_i+Q^kn_k\,(\delta_{ij}-5\,n_i n_j)
  \right].\nonumber
\end{eqnarray}
Here $n^i$ is a unit three-vector in the outward radial direction. The
coefficient $C$ of the spherically symmetric first term has been
normally taken to be zero in papers on the initial value problem on
maximal hypersurfaces. In this paper, we refer to $\tilde{\AW}_{ij}$
given above in Eq.~(\ref{eq:GeneralMomentumTensor}) as the
{\em generalized Bowen-York solution}.

Note that $\tilde{\AW}^{ij}$ is not invariant under the rescaling of
$R$ (see Eqs.~(\ref{eq:Rrescaling})). A scale-invariant effective source
term in Eq.~(\ref{eq:HamiltonianConstraint}) is 
\begin{equation}
  \label{eq:W} R^6 \tilde{\AW}_{ij} \tilde{\AW}^{ij} \equiv
  W(R,\theta,\varphi). 
\end{equation} 
The properties of $W$ are important for determining questions such as
the inversion symmetry of the hypersurface about the minimal
surface (see the Appendix). Also, the invariance of $W$ under the
rescaling freedom $R\to\Rrescale R$ implies that the parameters must
scale as 
\begin{equation}
  C\to C,\quad
  S^i\to S^i,\quad
  P^i\to\eta^{-1} P^i,\quad
  Q^i\to\eta Q^i.
\end{equation}

The physical interpretation of the Bowen-York parameters is not
necessarily the same on CMC hypersurfaces as on maximal hypersurfaces.
CMC hypersurfaces are not asymptotically flat. Identification of the
physical energy, linear momentum, and angular momentum of the system
on asymptotically null hypersurfaces is a non-trivial matter, in
general~\cite{Chrusciel2004}. In particular, the scaling dependence of
the boost parameters means that these cannot be interpreted as
physical momenta.

One can argue that in the limit of small $\TrKW$ and $C$ ($\TrKW M \ll
1$ and $C \sim (8M^2/3) (\TrKW M) \ll 1$), the geometry in the
vicinity of the black hole should be similar to that of solutions of
the zero mean curvature initial value problem found, for example, by
Cook~\cite{Thesis:Cook,cook90}. For hyperboloidal slices, the
conformal factor $\Omega$ is generally approximately constant at
intermediate distances, $\Omega M \ll R \ll \Omega/K$. (As can be 
seen in Fig.~\ref{fig:QuasiTrumpet}, $\Omega$ is
smaller near and inside the black hole, where $\Omega \sim R/M$, and
decreases toward zero approaching $\ScriPlus$, $\Omega < K R$.) These
intermediate distances are sufficiently close to the black hole that
the CMC slice still resembles a maximal slice, but far enough away to
be considered in the asymptotic regime. If $R$ is rescaled such that
$\Omega\approx 1$ in this intermediate regime, then the ADM formulas
for energy, linear momentum, and angular momentum should be at least
roughly valid. Therefore, one might identify the scaling invariant
$\Omega_{\rm max}P^i$ as a quasi-local linear momentum and $S^i$ as a
quasi-local angular momentum. However, these ``quasi-local'' values
may not match the correct physical values at future null infinity if
gravitational radiation is present outside the plateau region. As a
surrogate for the mass of the black hole, we use the ``irreducible
mass'' $\Mirr$, defined in the usual way from the area of the apparent
horizon $A_{\rm AH}$,
\begin{equation} \Mirr \equiv
  \left(\frac{A_{\rm AH}}{16\pi}\right)^{1/2}. 
\end{equation}
 
If the initial data is axisymmetric, then the physical angular
momentum can be calculated precisely using standard techniques. The
generalized Bowen-York solution for a single black hole is
axisymmetric provided that the boost is zero or aligned with the spin
vector. Then the coordinates can be chosen so that only the $z$
components of the spin and boost vectors are non-zero. The solution
for the conformal factor will be axisymmetric for our minimal surface
inner boundary condition, since the minimal surface is assumed to be a
coordinate sphere. Choose spherical polar coordinates ($R$, $\theta$,
$\varphi$) in the conformal flat space, so that the axial Killing
vector $\rightarrow{\xi} = {\partial}/{\partial\varphi}$. Then the
Komar angular momentum within a coordinate sphere is
\begin{align}\label{eq:QuasilocalSpin}
  J =& \; \frac{1}{8\pi}
  \int_0^\pi \int_0^{2\pi } {\alpha \sqrt{g} \, \xi^{R;t} d\theta
    d\varphi} \\ \nonumber =& -\frac{1}{8\pi} \int_0^\pi \int_0^{2\pi}
  {\tilde{\AW}_\varphi^R \, R^2 \sin(\theta)d\theta d\varphi} \\
  \nonumber =& \; S_z. 
\end{align}

In much of the earlier work on maximal hypersurfaces, considerable
emphasis has been placed on obtaining inversion-symmetric solutions of
the initial value problem (see~\cite{ksy83,Bowen-York:1980,cook91}).
Given a minimal surface at $R = R_{\rm ms}$, inversion symmetry
requires that $\Omega(R,\theta,\varphi) = (R_{\rm
  ms}/R)^2\,\Omega(R_{\rm ms}^2/R,\theta,\varphi)$. Solutions of
Eq.~(\ref{eq:HamiltonianConstraint}) for $\Omega$ with minimal surface
boundary conditions can be continued with inversion symmetry to $R <
R_{\rm ms}$ if and only if $W$ as defined in Eq.~(\ref{eq:W}) is
inversion symmetric. For $C=0$, the usual story is that the boost
terms are inversion-symmetric if $Q^i=\pm R_{\rm ms}^2 P^i$, with only
the minus sign applicable if spin is also present. We show in the
Appendix that with $C \ne 0$, inversion symmetry requires the {\em
  plus} sign when only boost is present, and that no inversion
symmetry is possible with both boost and spin unless the boost and
spin vectors are co-linear. Inversion symmetry is desirable primarily
for simplifying excision boundary conditions during evolution. On CMC
hypersurfaces, the generic absence of inversion symmetry requires
rethinking how excision will be handled, and it may be desirable to
just set $Q^i = 0$, as also done in~\cite{Brandt1997}.

The location of the apparent horizon relative to the minimal surface
is an important issue. The inner boundary of the computational domain
should be inside the apparent horizon. In spherical symmetry, the
apparent horizon will always be outside or on the minimal surface,
provided that a solution of the Hamiltonian constraint
exists. However, in the presence of a boost, the apparent horizon can
straddle the minimal surface~\cite{cook90}.  Care should be taken in
the choice of $R_{\rm ms}$ as the boost is increased, in order to
avoid this problem.

\subsection{Multi-black hole solutions}
\label{Sec:MultiBhIVP}
The goal of this section is to construct a CMC hypersurface with $N$
black holes, with masses (approximately) $M_\alpha$, Bowen-York
parameters $\vec P_\alpha$ and $\vec S_\alpha$ at coordinate locations
$\vec c_\alpha$.  Here $\alpha$ labels the black holes.

The setup for single black holes in Sec.~\ref{Sec:SchwarzIVP} can be
generalized by choosing one excision boundary for each black hole,
with radius $R_{\alpha}$ centered at $\vec c_\alpha$.  Because the
momentum constraint is linear, the extrinsic curvature can be taken as
the superposition of $N$ copies of
Eq.~(\ref{eq:GeneralMomentumTensor}), each one centered at the
appropriate $\vec c_\alpha$:
\begin{equation}
  \tilde{\AW}_{ij} = \sum_{\alpha} \left(\tilde{\AW}_{ij}^{\alpha\,C} +
  \tilde{\AW}_{ij}^{\alpha\,P} + \tilde{\AW}_{ij}^{\alpha\,Q} +
  \tilde{\AW}_{ij}^{\alpha\,S}\right)
\end{equation}
If the black holes are sufficiently widely separated, and if the outer
boundary is sufficiently far away, we expect that close to each of
these black holes, the solution is a perturbation of the single black
hole case.

In the asymptotically flat case, the conformal factor is close
  to unity, except very close to each black hole.  Therefore, the
  coordinate distance $|\vec c_\alpha - \vec c_{\beta}|$ between the
  black holes $\alpha$ and $\beta$ is a convenient and reasonably
  accurate approximation of the proper separation between the black
  hole horizons.  Because of the rescaling freedom discussed in
  Eqs.~(\ref{eq:Rrescaling}), $\Omega$ may not be close to unity on
  the hyperboloidal slices considered here, and therefore the
  coordinate distance may deviate significantly from the proper
  separation.  

  There is only {\em one} global value $\TrKW$ and {\em one} value
  $R_+$ for the whole multi black-hole configuration, whereas each
  black hole has its ``own'' constants $C_\alpha$, $R_\alpha$, as well
  as $P^i_\alpha$ and $S^i_\alpha$. Therefore, the interesting
  question arises of how to use $C_\alpha$ and $R_\alpha$ to control
  properties of the individual black holes, for instance their masses,
  given {\em fixed} values for $\TrKW$ and $R_+$. Assuming that the
  presence of $P^i_\alpha$ and $S^i_\alpha$ will only mildly perturb
  the case of the spherically symmetric black hole, we can use
  Fig.~\ref{fig:ContoursKM} to address this question. A given $\TrKW$
  and a (desired) value for $M=M_\alpha$ places the solution on a
  particular $\TrKW M=$ constant contour. Given a desired value for
  $\gamma\equiv r_{\rm ms}/r_H$, a unique point in this figure is
  determined, and one can read off $R_{\rm ms}/R_+$ and $\TrKW
  C^{1/2}$ and then compute $R_\alpha=R_{\rm ms}$ and $C_\alpha=C$.

\begin{figure}
\includegraphics[width=0.9\linewidth]{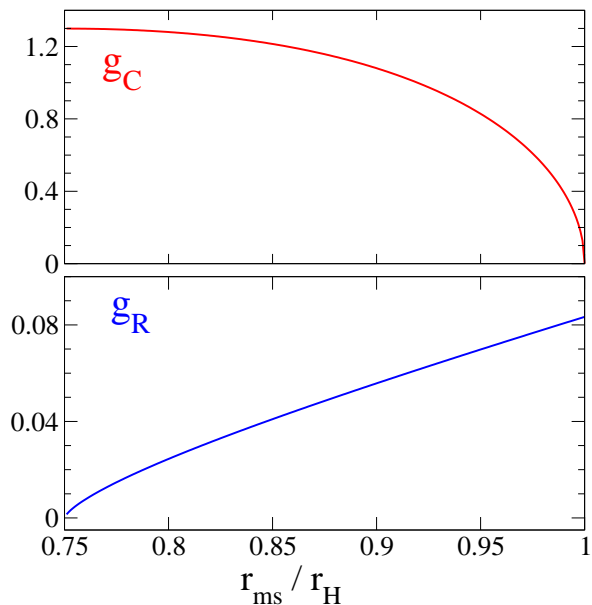}
\caption{\label{fig:AsymptoticBehavior}The functions $g_C$ and $g_R$,
  which are the asymptotic values of $C/M^2$ and
  $R_{\rm ms}/(R_+ \; \TrKW M)$ in the limit $\TrKW M\to 0$.  These
  functions are useful for choosing excision radius and $C$ when
  constructing CMC initial data for black holes with given $\TrKW$, $M$
  and outer boundary radius $R_+$.  }
\end{figure}

  This procedure can be simplified in the particularly interesting
limit $\KW M\ll 1$. Consider a minimal surface with $3/4<r_{\rm
ms}/r_H\equiv\gamma<1$, and with a given value of $\TrKW M$.
Substituting $\TrKW M$ and $\TrKW r_{\rm ms}=2\gamma K M$ into
Eq.~(\ref{eq:f=zero}), one can solve for $\TrKW^2 C$ and compute $C$.
The result is
\begin{equation}\label{eq:gC}
C\approx g_C(\gamma)\, M^2,\qquad \TrKW M\ll1
\end{equation}
with $g_C(\gamma)=4\gamma\sqrt{\gamma-\gamma^2}$. 
We also find numerically that
\begin{equation}\label{eq:gR}
\frac{R_{\rm ms}}{R_+}\approx g_R(\gamma)\,\TrKW M,\qquad \TrKW M\ll 1.
\end{equation}
Therefore, if one knows the coefficients $g_C$ and $g_R$ for the
desired ratio $\gamma=r_{\rm ms}/r_H$, one can immediately compute the
values for $C$ and $R_{\rm ms}/R_+$ from Eqs.~(\ref{eq:gC})
and~(\ref{eq:gR}). These coefficients $g_C(\gamma)$ and $g_R(\gamma)$
are plotted in Fig.~(\ref{fig:AsymptoticBehavior}).

\subsection{Trumpet Inner Boundary}
\label{Sec:Trumpet}

In the absence of spherical symmetry, a trumpet solution of the
Hamiltonian constraint equation can be understood as the limit $R_{\rm
  ms} \to 0$. That is, the Hamiltonian constraint
Eq.~(\ref{eq:HamiltonianConstraint}) is solved with the boundary
condition $\partial{\Omega}/\partial{R} = \Omega/R$ at $R=0$. The
consequence of this boundary condition is that $\Omega = 0$ and
$\Omega/R$ is finite at $R=0$, which in turn means that the proper
distance from finite $R$ to $R=0$, $\int_{0}^{R}d R/\Omega$, is
infinite. In this section, we show how the singularities in the
equation determine the non-singular solution at the trumpet inner
boundary and derive the behavior of the solution close to the trumpet
boundary.

Note that a necessary condition for a non-singular solution is that
the second ``boost'' vector $Q^i = 0$. Otherwise, the $W$ source term
in the Hamiltonian constraint blows up at $R = 0$. This is a very
reasonable condition which follows automatically from inversion
symmetry in the limit $R_{\rm ms} \to 0$ (see the Appendix), and
reflects the fact that the other side of the Einstein-Rosen bridge is
infinitely far away from any point with $R > 0$.

We begin by rewriting Eq.~(\ref{eq:HamiltonianConstraint}) with $U
\equiv \Omega/R$ as the dependent variable. The new form of the
equation is
\begin{align}
\label{Eq:Trumpet2}
  &R^2\frac{\partial^2 U}{\partial R^2} + {4}R\frac{\partial U}
      {\partial R} + {2}U + \hat \Delta U  = \nonumber \\
  &\frac{3}{{2}U}\left[ {\left( {U + R\frac{\partial U}{\partial R}}
      \right)^2 + \hat \nabla U \cdot \hat \nabla U - \left( {\frac{\TrKW}
          {3}} \right)^2 + \frac{1} {6}U^6 W } \right],
\end{align}
where $\hat \Delta$ is the Laplacian operator and $\hat \nabla$ the
gradient operator on the unit two-sphere, and $W$ is defined by
Eq.~(\ref{eq:W}).  We have assumed a conformally flat spatial metric.

Now let $U = U_0(\theta,\varphi) + R^{\alpha} U_1(\theta,\varphi) +
\ldots$ and $W = W_0(\theta,\varphi) + R W_1(\theta,\varphi) +
\ldots$. From the expression for $W$ in Eq.~(\ref{eq:AtildeSquared}),
we see that
\begin{equation}
  \label{Eq:W0def}
  W_0 = {6}(C^2 + 3\sin^2(\psi)\; S^iS_i)
\end{equation}
for any boost $P^i$, where $\psi$ is the angle with the spin
direction, equal to $\theta$ if the spin is along the polar axis.
Unless the boost is non-zero, $W = W_0$ at all R and $W_1 = 0$.
 
In zeroth order, we get
\begin{equation}
\label{Eq:Trumpet4}
\hat \Delta U_0 = - \frac{1}{2}U_0 + \frac{3}{{2U_0}}\left[ {\hat \nabla U_0
    \cdot \hat \nabla U_0 - \left( {\frac{\TrKW}{3}} \right)^2 + W_0
    U_0^6 } \right],
\end{equation}
which has a unique solution regular everywhere on the unit sphere for
any $\TrKW > 0$, any value of $C>0$, and any spin vector $S^i$.
Uniqueness can be demonstrated using a method of Moncrief
\cite{Moncrief86} applied to the quasi-linear form of the equation
obtained by the change of variable $U_0 \to 1/V^2$.

In the absence of spin, $U_0$ is independent of angle and $U_0^2$
  is the solution of the cubic equation
\begin{equation}
\label{Eq:Trumpet5}
C^2 \left({U_0^2}\right)^3 - \frac{1}{3}U_0^2 - \left({\frac{\TrKW}{3}} \right)^2 = 0.
\end{equation}
The only positive real root if $3 \TrKW^2 C/2 \le 1$ is
\begin{equation}
\label{Eq:Trumpet6}
U_0^2 = \frac{2}{{3C}}\cos \left[ {\frac{1}{3}\cos ^{ - 1} \left(
    {\frac{3}{2}\TrKW^2 C} \right)} \right].
\end{equation}
The trigonometric functions are replaced by the corresponding
hyperbolic functions if $3 \TrKW^2 C/2 \ge 1$.

The next-to-leading terms in Eq.~(\ref{Eq:Trumpet2}) give an equation 
for $U_1$:
\begin{align}
\label{Eq:Trumpet7}
&\hat \Delta U_1 - 3\frac{\hat \nabla U_0 \cdot \hat \nabla U_1}{U_0} \\ \nonumber
&+ \left( {\alpha ^2 + 1 - \frac{3}{2}W_0 U_0^4 +  \frac{\hat \Delta U_0}{U_0}}
\right) U_1 \\ \nonumber
& =  \frac{3}{2} U_0^5 W_1 R^{1-\alpha}.
\end{align}
If $W_1 \not= 0$, the solution of the inhomogeneous
equation with $\alpha = 1$ gives the leading contribution to $U_1$. 
There is also a unique lowest value of $\alpha$ for which the 
{\it homogeneous} equation has a non-trivial solution regular 
everywhere on the unit sphere. This solution to the homogeneous 
equation, times $R^{\alpha}$,
will, with a coefficient undetermined by the trumpet boundary
condition, contribute to $U-U_0$. The coefficient is fixed by the
requirement that the global solution for $\Omega$ satisfy the $\Omega
= 0$ boundary condition at future null infinity.

If the spin is zero, $W_0 = {6}C^2$ and the homogeneous 
$\alpha$ is the solution of the algebraic equation
\begin{equation}
\label{Eq:Trumpet8}
\alpha ^2 + 1 = {9}C^2 U_0^4 = 
{4}\cos^2 \left[ {\frac{1}{3}\cos ^{ - 1}
    \left( {\frac{3}{2}\TrKW^2 C} \right)} \right].
\end{equation}
In the range $0 \le 3 \TrKW^2C/2\le 1$ of most interest, Eq.
(\ref{Eq:Trumpet8}) implies $\sqrt{2} \le \alpha \le \sqrt{3}$. For
larger $\TrKW^2C$, the trigonometric functions are replaced by
hyperbolic functions and $\alpha$ continues to increase.

In practical terms, there is very little difference between a
  solution satisfying the exact trumpet boundary condition and a
  solution satisfying the minimal surface boundary condition with a
  very small, but non-zero, $R_{\rm ms}$.  Very small means that
  $R_{\rm ms}/R_{\rm AH} \ll 1$. For the Schwarzschild case, $R_{\rm
    ms}/R_+$ should be far below the heavy black line in
  Fig.~\ref{fig:ContoursKM}.

Finally, all of this discussion has been in the context of single black 
holes.  With multiple black holes, each trumpet boundary must be 
treated separately and matched to the global solution on a surface 
surrounding the black hole.  The analysis right at the trumpet 
boundary is not affected by the presence of other black holes, since 
the $R^6$ factor in $W$ kills the finite contribution of the other black 
holes to the conformal traceless extrinsic curvature at $R=0$.

%%%%%%%%%%%%%%%%%%%%%%%%%%%%%%%%%%%%%%%%%%%%%%%%%%%%%%%%%%%%%%%%
\section{Numerical Results}
\label{Sec:3}
%%%%%%%%%%%%%%%%%%%%%%%%%%%%%%%%%%%%%%%%%%%%%%%%%%%%%%%%%%%%%%%%

In this section, we numerically construct a variety of hyperboloidal
initial data sets using the generalized Bowen-York solution. These
results are obtained with a pseudo-spectral elliptic solver that is
part of the Spectral Einstein Code, {\tt SpEC}. This solver is
described in detail in Ref.~\cite{Pfeiffer2003}. The desired solution
is expanded in terms of spherical harmonics and Chebyshev
polynomials. Truncation at some finite expansion order results in an
algebraic system of equations for the expansion coefficients or,
equivalently, for the values of the solution at the collocation
points. This system is solved with a Newton-Raphson technique,
employing the preconditioned generalized minimal residual method
(GMRES)~\cite{Saad:1993} to solve the linearized system of equations
at each iteration using the software package
PETSc~\cite{petsc-web-page,petsc-user-ref,petsc_efficient}.  The SpEC
elliptic solver has been used on a wide variety of formulations of the
initial value problem (see, e.g.~\cite{Pfeiffer2002a,Pfeiffer2004}),
including puncture initial data ~\cite{Dennison2006,Lovelace2008}
(which also uses the Bowen-York extrinsic curvature). Below, we
present results of convergence tests of our initial data.

\subsection{Spherical Symmetry}
\label{Sec:NumResults:Schwarz}

\begin{figure}
  \includegraphics[width=0.9\columnwidth]{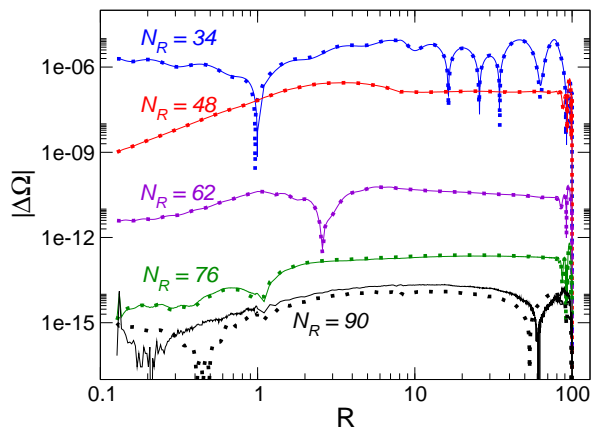}
  \caption{\label{fig:ConvergenceSchwarzschild} Convergence of the
    numerical solution of the Hamiltonian constraint for a
    Schwarzschild black hole. Plotted are five different resolutions
    $N_R$, where $N_R$ is the number of radial collocation points.
    Solid lines show the difference from the analytic solution, dotted lines
    the difference from the numerical solution at the next higher
    resolution. The highest resolution has $N_R=104$.}
\end{figure}

As a first test, we reproduce the analytically known spherically
symmetric solutions discussed in Sec.~\ref{Sec:Schwarzschild}, using
the numerical approach described in Sec.~\ref{SubSec:SpherSym}. We
choose $M=0.85$, $\TrKW=0.1$, $r_{\rm ms}/r_H=0.8$ and $R_+=100$. The
relations shown in Fig.~\ref{fig:ContoursKM} then imply $C=1.0086$ and
$R_{\rm ms} = 0.127$. From these numbers, only $K$, $C$, $R_{\rm
  ms}=R_1$, and $R_+=R_2$ are used in the numerical solution; the
other numbers are used only when computing the analytical solution
with which to compare. Fig.~\ref{fig:ConvergenceSchwarzschild} shows
convergence of the numerical solution to the analytic solution. The
solid lines plot the differences between the numerically determined
$\Omega$ and the analytic solution of Eq.~(\ref{eq:RmsOverRplus-u})
computed with {\tt Mathematica}. As we increase the resolution of the
elliptic solver, we find exponential convergence to the analytic
solution. For the generic examples considered later in this paper
(which include spin, boost, and two black holes), no analytic
solutions are known. Therefore, in
Fig.~\ref{fig:ConvergenceSchwarzschild}, we also present an estimate
of the numerical error which does not rely on knowledge of the
analytic solution. Specifically, the dotted lines show the differences
between the numeric solutions at two successive resolutions. As can be
seen, these track very closely the error obtained from comparing the
lower resolution run to the analytic solution. 

In our second example, we explore a solution which is very close to
the trumpet configuration. Recall that for a trumpet, for $R$ close to
$R=0$, the solution behaves as $\Omega=U_0R$ with $U_0$ given in
Eq.~(\ref{Eq:Trumpet6}). We choose parameters $C=1$, $K=0.1$, $R_{\rm
  ms}=10^{-6}$, and $R_+=100$ which, as can be seen from
Fig.~\ref{fig:ContoursKM}, result in an inner boundary which is very
close to the trumpet limit. Because application of the minimal surface
condition in this case proved numerically problematic (presumably due
to dividing by the very small number $R_{\rm ms}$), we use the
Dirichlet condition $\Omega=U_0 R_{\rm ms}$ at the inner boundary.

\begin{figure} \includegraphics[width=0.8\columnwidth]{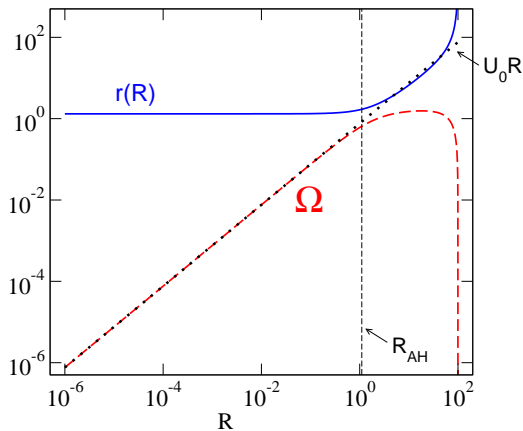}
  \caption{\label{fig:QuasiTrumpet} Numerical solution for a
    Schwarzschild black hole with the inner boundary very close to the
    trumpet limit. The dashed red line shows the conformal factor
    $\Omega$, and the solid blue line shows the Schwarzschild radius
    (calculated from the proper area of coordinate spheres) as a
    function of the conformal radius. The dotted black line is $U_0R$,
    which equals $\Omega$ for a trumpet, with $U_0$ given in
    Eq.~(\ref{Eq:Trumpet6}). The vertical dashed line locates the
    apparent horizon.}
 \end{figure}

The numerical solution of the Hamiltonian constraint equation for this
example is shown in Fig.~\ref{fig:QuasiTrumpet}. One sees that in the
region $10^{-6} \le R \le 1$, the conformal factor $\Omega$ is
proportional to $R$, with $U_0$ the constant of proportionality.
Furthermore, within this range of conformal radius, the proper area of
coordinate spheres ($4\pi r^2$, where $r$ is the Schwarzschild radius) is
approximately constant, which is consistent with the long cylinder of
the trumpet. The location of the apparent horizon $r_H=2M=1.69$ is
shown as a vertical dashed line in this figure.

\subsection{Single spinning black hole}
\label{Sec:NumResults:SingleSpin}

Here, we construct a single spinning black hole, with no boosts. We
take the Bowen-York spin parameter $S^i=(0,0,S)$ and solve the
Hamiltonian constraint for the conformal factor $\Omega$, with varying
$S$. The solution is axisymmetric, so $S^i$ represents the total
angular momentum of the black hole (see
Sec.~\ref{SubSec:BhPlusSpinBoost}).

\begin{figure}
  \includegraphics[width=0.8\columnwidth]{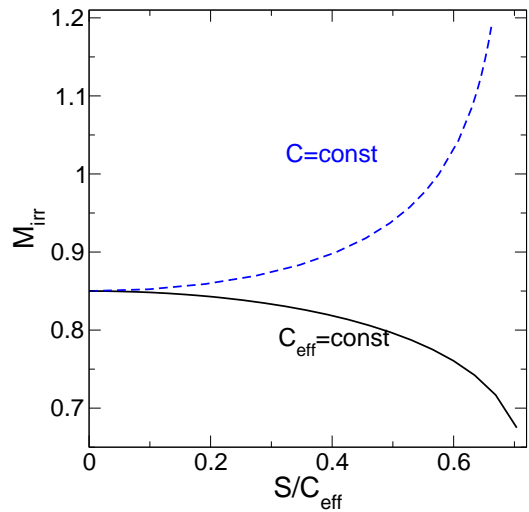}
  \caption{\label{fig:MirrVsSOverCeff} Irreducible mass versus
    dimensionless spin for a single black hole spinning around the
    z-axis.}
\end{figure}

In the absence of boosts, $W$ (Eq.~(\ref{eq:W})) reduces to $W_0$ as
defined by Eq.~(\ref{Eq:W0def}), with $\psi = \theta$. The radial
behavior of $\Omega$ will be rather similar to the spherically
symmetric solution with the same parameters as long as the spherically
symmetric $C^2$ is replaced by the solid angle average of $W/6$, which
we denote by 
\begin{equation}
\label{eq:Ceff} 
C_{\rm eff}^2 \equiv C^2 + {2}S^2. 
\end{equation} 
We use $S/C_{\rm eff}$ as a dimensionless measure of the importance of
spin. Note that $0 \leq S/C_{\rm eff} < 1/\sqrt{2}$ as $S/C$ varies
from zero to infinity. We find that the irreducible mass varies less
with spin keeping $C_{\rm eff}$ constant than when keeping $C$
constant, particularly for large spins, as shown in
Fig.~\ref{fig:MirrVsSOverCeff}. The constant values of $C$ and $C_{\rm
  eff}$ are 1.0086, with $K=0.1$, $R_{\rm ms}=0.127$, and $R_+=100$.

\begin{figure}
  \includegraphics[width=0.9\columnwidth]{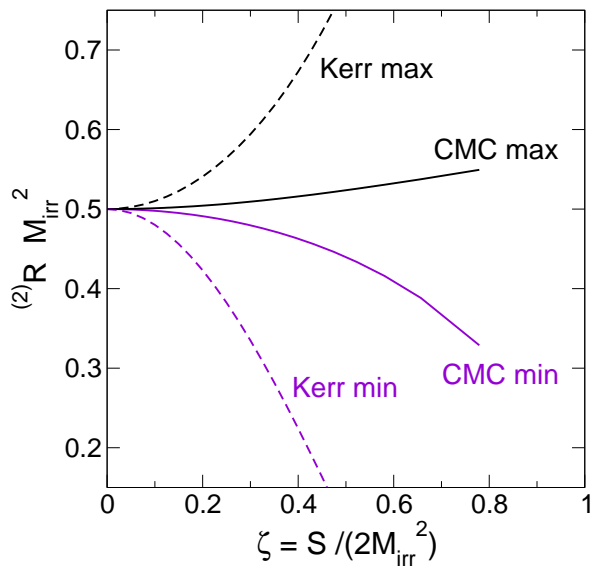}
  \caption{\label{fig:MinMax2RicciScalar} Dimensionless scalar
    curvature of the apparent horizon versus spin for a single black
    hole. The solid lines are the maximum and minimum numerical values
    computed on CMC slices. The dashed lines are the maximum and
    minimum analytic values for a Kerr black hole (these lines
    continue to the maximal Kerr value $\zeta=1$, where the maximum
    and minimum are $2$ and $-1/2$, respectively).}
\end{figure}

In Fig.~\ref{fig:MinMax2RicciScalar}, we study how the non-zero spin
distorts the intrinsic geometry of the apparent horizon and compare
the results to an analogous distortion computed from the analytic Kerr
solution. The solid lines of Fig.~\ref{fig:MinMax2RicciScalar} show
the maximum and minimum of the Ricci scalar $^{(2)}{\cal{R}} \Mirr^2$
computed from the 2-metric induced on the apparent horizon of the
spinning black hole. The dashed lines show the maximum and minimum
values calculated from the analytic Kerr solution, taken from Eq.~(B1)
of Ref.~\cite{Lovelace2008}. Note that deviations from $0.5$ are
deviations from a spherical geometry. We see that the apparent horizon
distortion is much less for our conformally flat initial data than it
is for Kerr. The CMC data plotted in Fig.~\ref{fig:MinMax2RicciScalar}
is the same shown in Fig.~\ref{fig:MirrVsSOverCeff}, with the CMC
curves terminating at ${\rm max}(S/C_{\rm eff})=1/\sqrt{2}$.

The horizontal axis in Fig.~\ref{fig:MinMax2RicciScalar} is the 
spin-extremality parameter 
\begin{equation}
\zeta=\frac{S}{2M_{\rm irr}^2}
\end{equation}
as introduced in Ref.~\cite{Lovelace2008}. A maximally spinning Kerr
black hole has $\zeta=1$, and the CMC-sequence considered here allows
values as large as $\zeta\approx 0.78$. In Sec.~\ref{Sec:Discussion},
we shall place this number into the context of results on zero mean
curvature slices.

\subsection{Single boosted black hole}
\label{Sec:NumResults:SingleBoost}

Next, we construct single, non-spinning, boosted black holes. We shall
vary $P^i$ and shall choose $Q^i=+R_{\rm ms}^2P^i$, in order to make
the black hole spacetime inversion symmetric (see the Appendix for
details). As we vary the boost, we keep the irreducible mass of the
constructed black holes constant by a suitable choice of $R_{\rm ms}$.
Specifically, $M_{\rm irr}\!=\!0.85$ and the remaining CMC parameters
are chosen to be $R_{\rm +}\!=\!100$,
  $\TrKW\!=\!0.1$, and $C\!=\!1.0086$. 

\begin{figure}
  \includegraphics[width=0.8\columnwidth]{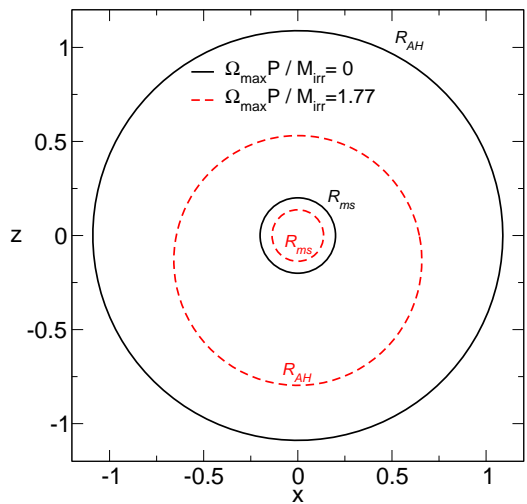}
  \caption{\label{fig:AH_L08_xz_Pz1VsPz0} Coordinate locations of the
    apparent horizons and minimal surfaces, cut through the x-z plane,
    for a non-spinning, unboosted black hole (solid black circles),
    and for a non-spinning black hole boosted in the z-direction
    (dashed red circles).}
\end{figure}

  First, we compare initial data sets for an unboosted black hole and
  for a boosted black hole with $P \Omega_{\rm max}/\Mirr=1.77$.
  Fig.~\ref{fig:AH_L08_xz_Pz1VsPz0} shows the coordinate locations of
  both the apparent horizon and the minimal surface for these two
  cases. The apparent horizon remains an approximate coordinate
  sphere, although its coordinate radius is reduced (recall that
  $M_{\rm irr}$ is identical for the boosted and unboosted data set,
  which was achieved by reducing $R_{\rm ms}$ for the boosted case).
  Furthermore, the apparent horizon is offset from the excision sphere
  in a direction opposite to the boost $P^i$, analogous to the
  behavior of asymptotically flat inversion symmetric Bowen-York
  initial data~\cite{cook90}.

\begin{figure}
 \includegraphics[width=0.8\columnwidth]{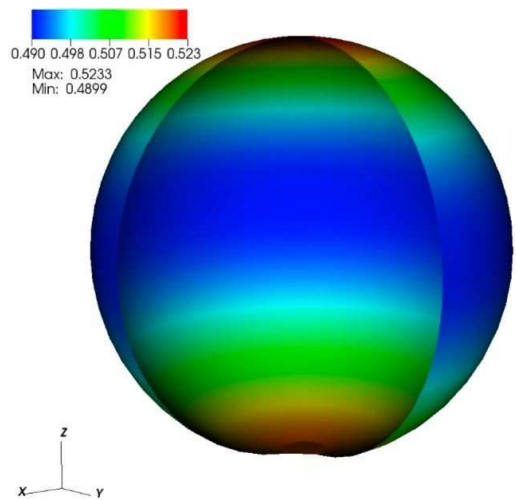}
 \caption{\label{fig:RicciForBoostedBH} 2D Ricci scalar on the
   apparent horizon surface of a black hole with 
   $\Omega_{\rm max} P/\Mirr=1.77$. In this view, a
wedge-shaped region has been removed from the front.}
\end{figure}

To investigate the intrinsic geometry of the apparent horizon for the
boosted black hole, we compute the Ricci scalar $^{(2)}{\cal{R}}
\Mirr^2$ from the 2-metric induced on the apparent horizon.
Fig.~\ref{fig:RicciForBoostedBH} plots this quantity; it is
axisymmetric (as it must be), is maximum at the poles along the z-axis
and minimum along the equatorial region. (Recall that $^{(2)}{\cal{R}}
\Mirr^2=0.5$ for a spherical geometry.) The numerically computed spin
of this black hole is indeed zero, to machine precision.

\begin{figure}
 \includegraphics[width=0.9\columnwidth]{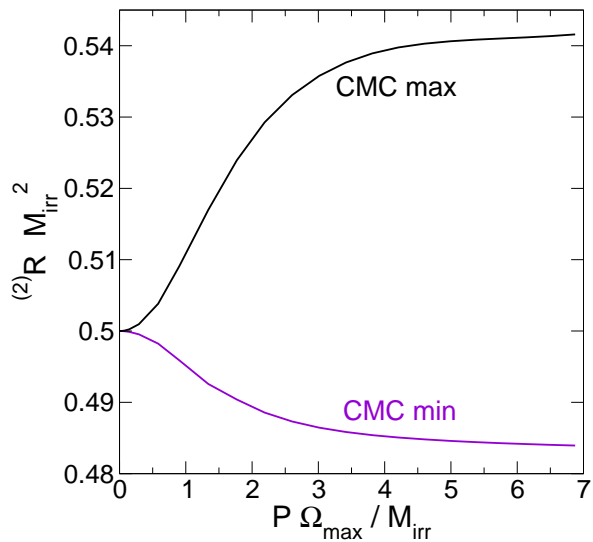}
 \caption{\label{fig:DimlessPVs2DRicciMinMax} Dimensionless intrinsic
   geometry of the apparent horizon (2D Ricci scalar, times the
   irreducible mass squared) versus $P\Omega_{\rm max}/\Mirr$
   for a single non-spinning black
   hole, with the ratio
   $R_{\rm AH}/R_{\rm ms}$ kept fairly constant. Shown are the
   maximum and minimum numerical values computed on CMC slices.}
\end{figure}

Fig.~\ref{fig:DimlessPVs2DRicciMinMax} shows the minimum and maximum
of $^{2}{\cal{R}} \Mirr^2$ as the boost parameter is varied in the
range $0 \le \Omega_{\rm max} P/\Mirr \le 6.87$. The minimum and
maximum values of $^{2}{\cal{R}}_{\rm max} M_{\rm irr}^2$ when
$\Omega_{\rm max} P/\Mirr=1.77$ agree with those shown in
Fig.~\ref{fig:RicciForBoostedBH}.

\subsection{Binary black hole initial data} 
\label{Sec:NumResults:BBH}

To demonstrate the generality of the approach that we have presented,
we shall construct initial data for two black holes with mass-ratio
approximately $2:1$ and non-zero, arbitrarily oriented Bowen-York spin
and boost parameters. First we describe how we obtain input parameters
for the elliptic solver corresponding to our particular physical
parameters. First, we choose $\gamma \equiv r_{\rm
  ms}/r_H=0.8$, which singles out a particular line of constant
$\gamma$ in Fig. \ref{fig:ContoursKM}, for each black hole. Next, we
pick $\TrKW M$ for each black hole so that (i) its minimal surface is
at least partially down the throat of the trumpet, which is near the
turnover of the $\gamma=0.8$ curve and (ii) Eqs. (\ref{eq:gC}) and
(\ref{eq:gR}) hold, i.e. before the turnover. With these criteria in
mind, we choose $\TrKW$ (a global parameter) to be $0.05$ and the
masses of black holes A and B to be, respectively, $M_A = 2/3$ and
$M_B=1/3$. Finally, from Eq.~(\ref{eq:gR}) and
Fig.~\ref{fig:AsymptoticBehavior}, we find $R_{\rm
  ms}/R_+=8.1\times 10^{-4}$ for hole A and $R_{\rm ms}/R_+=4.1\times
10^{-4}$ for hole B.

We fix the overall length scale by setting $R_+=300$. This places the
excision radii at $R_{\rm ms}=0.244$ and $0.122$ for holes A and B,
respectively, and the apparent horizon radii $R_{\rm AH}\approx 1$
(because from Fig.~\ref{fig:ContoursKM}, $R_{\rm ms}/R_{\rm AH}\approx
0.2$). The coordinate locations of the two holes are then chosen to be
$(x_A,y_A,z_A)=(10,0,0)$ and $(x_B,y_B,z_B)=(-20,0,0)$, and the center
of mass of the holes is at the origin of the coordinate system.

We take the spins to be $S_A^i=(0,0,S_A)$ and $S_B^i=(S_B,0,0)$. Since
we are adding significant spins, it is necessary to set $C_{\rm
  eff}$ (defined in Eq.~(\ref{eq:Ceff})) equal to $g_C M^2$ for each
hole, giving $0.569$ for hole A and $0.142$ for hole B. We take
$S_A=0.4$ and $S_B=0.1$, then giving $C_A=0.0613$ and $C_B=0.0128$.

We take the boost parameters of the two black holes to be equal and
opposite in the $y$-direction, with magnitude $P_A=P_B=0.067$. This
gives approximate speeds of $v_A=\Omega_{\rm max} P_A/M_A=0.24$ and
$v_B=\Omega_{\rm max} P_B/M_B=0.48$, with $M_A=2/3$ and $M_B=1/3$ as
given above. With $C_A$ and $C_B$ not equal to zero, and $P^i$ not
co-linear with $S^i$, inversion symmetry is not possible (refer to the
Appendix). Thus, we set $Q_A^i=Q_B^i=0$.

\begin{figure}
  \includegraphics[width=0.9\columnwidth]{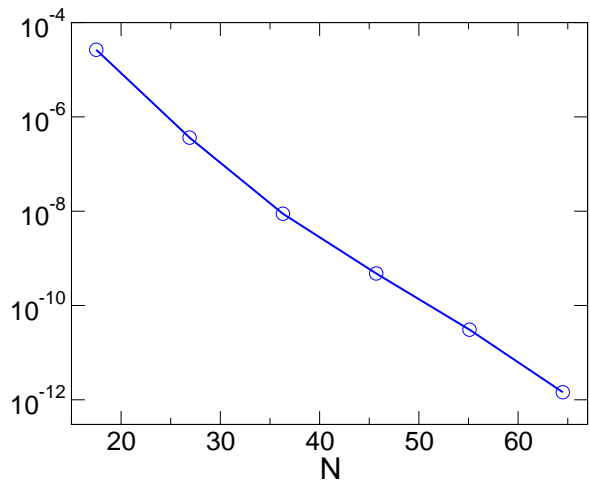}
  \caption{\label{fig:ConvergenceBBH} Convergence of the elliptic
    solver for the unequal mass binary black hole example shown in
    Fig.~\ref{fig:BBH}. Shown is the volume L2-norm of the residual of
    $\Omega$ as $N$ is increased, where $N$ is the cube root of the
    total number of collocation points.}
\end{figure}

Fig.~\ref{fig:ConvergenceBBH} shows exponential convergence of the
volume L2-norm of the residual for the solution $\Omega$ of the
elliptic solver in this example, as the resolution of the numerical
grid is increased. In addition, we have calculated the irreducible
masses of the two holes and find values of $0.53$ for
hole A and $0.27$ for hole B. This gives a mass ratio of $1.96$.

\begin{figure} \includegraphics[width=0.8\columnwidth]{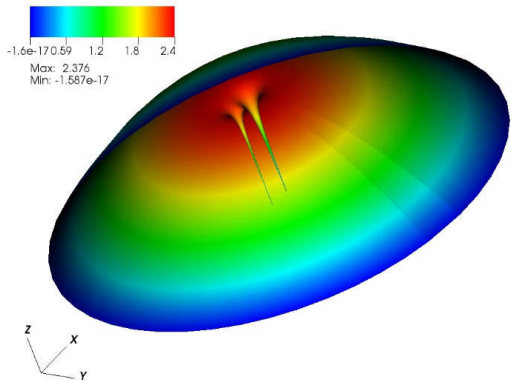}
  \caption{\label{fig:BBH} Conformal factor $\Omega$ for a spinning,
    boosted binary black hole, with mass ratio 2:1. 
    The inner boundaries are the minimal surfaces of the two black
    holes, and the outer boundary is null infinity. The maximum of
    $\Omega$ is red, and equals 2.38. The minimum is dark blue, which
    is zero to machine precision.}
\end{figure}

Fig.~\ref{fig:BBH} shows the conformal factor on the full
computational domain for the mass ratio $2:1$ boosted, spinning binary
black holes described above. The dark blue color at the outer edge
shows that $\Omega=0$ at null infinity (to machine precision). In the
middle, there is a prong-like feature, the tips of which are the two
black holes. It is evident that the conformal factor becomes quite
small in the vicinity of the two black holes.

Since our calculation of input parameters assumes spherical symmetry
when our holes in fact have appreciable spins and boosts, one expects
the irreducible masses to differ somewhat from the values used for
calculating the input parameters. This is indeed what we find ($0.53$
vs. $0.67$ for hole A and $0.27$ vs.  $0.33$ for hole B). Finally, we
find that the intrinsic geometry of each hole is distorted by the same
amount. In particular, the minimum and maximum values of
$^{(2)}{\cal{R}} \Mirr^2$ are, respectively, $0.37$ and $0.54$ for
each hole.

% For A, $0.315<R_{\rm AH}<0.327$. For B, $0.151<R_{\rm AH}<0.161$.

%%%%%%%%%%%%%%%%%%%%%%%%%%
\section{Discussion}
\label{Sec:Discussion}
%%%%%%%%%%%%%%%%%%%%%%%%%%

In this paper, we have considered the conformal method on CMC
hyperboloidal slices, focusing on generalizing the traditional
Bowen-York data. There are two key aspects that make Bowen-York data
easy to construct. First, for {\em constant} mean curvature (no matter
whether $\TrKW=0$ or $\TrKW\neq 0$), the momentum constraint decouples
from the Hamiltonian constraint, and the former can be solved first.
Second, with {\em conformal flatness}, the momentum constraint
simplifies to such an extent that analytical solutions are known: the
symmetric, tracefree, divergence-free tensors with appropriate radial
fall-off.  Interestingly, the second aspect carries over from zero
mean curvature to non-zero mean curvature, with the conformal factor
now playing the dual role of turning the Hamiltonian constraint into
an elliptic equation as well as compactifying $\ScriPlus$. The general
Bowen-York conformal traceless extrinsic curvature still solves the
momentum constraint analytically, where the only change necessary is
addition of a spherically symmetric divergence-free tensor [the first
term in Eq.~(\ref{eq:GeneralMomentumTensor})].

Compared to the zero mean curvature case as usually formulated, the
hyperboloidal initial value problem has more free parameters, most
notably the constants $\TrKW$ (the mean curvature) and $C$ (the
coefficient of the spherically symmetric contribution to the conformal
traceless extrinsic curvature), though a non-zero $C$ is also
consistent with zero mean curvature. Both these constants, as well as
the minimal surface coordinate radius, have to be chosen carefully,
and a significant portion of Sec.~\ref{Sec:2} is devoted to working
out permissible choices, and their consequences on the initial data
under construction.

As in the zero mean curvature case, hyperboloidal Bowen-York initial
data trivially extends to multiple black holes with different spin-
and boost-parameters for each black hole. Once again, one must be
careful to choose the constants $C$ (one for each black hole) and the
radii of the excision boundaries, and Sec.~\ref{Sec:MultiBhIVP} gives
simple rules how to do this. However, it is worth noting one
significant difference. A single black hole {\it must} be centered at
the origin of the conformal coordinates on a CMC hypersurface to be
precisely Schwarzschild, since the outer boundary condition is imposed
at a finite coordinate radius. A displaced black hole is not
spherically symmetric.

For hyperboloidal slices, the elliptic equation for the Hamiltonian
constraint, Eq.~(\ref{eq:HamiltonianConstraint}), is singular at the
outer boundary $\ScriPlus$, where $\Omega\to 0$. Perhaps surprisingly,
we have not encountered any difficulties when numerically solving this
equation, for either single or binary black hole initial data. This is
without any attempt to isolate and explicitly cancel the singular
terms in the equation at future null infinity, as was advocated
in~\cite{FrauendienerLRR}. We suspect that the absence of numerical
difficulties is related to the simple Dirichlet boundary condition
$\left.\Omega\right|_{\ScriPlus}=0$, and to the fact that the singular
terms force the solution to also satisfy the von Neumann condition
$(\partial\Omega/\partial R)_{\ScriPlus} = -\TrKW/3$, which follows
from Eq.~(\ref{eq:OmegaOnScriPlusII}), implying spherical symmetry to
at least first order in an expansion away from null infinity. The
freedom in the solution at the outer boundary necessary to accommodate
a global solution of the elliptic equation also satisfying an inner
boundary condition at $R_{\rm ms}$ resides in a higher order term in
the expansion of $\Omega$ away from $\ScriPlus$. Our spectral code
never evaluates the Hamiltonian constraint
Eq.~(\ref{eq:HamiltonianConstraint}) right {\em at}~ $\ScriPlus$.
Rinne~\cite{Rinne2009} has also had no difficulty in solving
the same elliptic equation with a finite difference code, as part of a
constrained evolution scheme on CMC hypersurfaces \cite{Moncrief2009}.

The Hamiltonian constraint equation is also singular at the inner
boundary in the special case of a trumpet, for which $\Omega = 0$ at
$R=0$.  This is a more challenging numerical problem, as discussed in
Sec.~\ref{Sec:Trumpet}.  Singular terms include some inside the
Laplacian operator.  Their cancellation again uniquely determines the
normal derivative of $\Omega$ there, but now that will have angular
dependence if the spin is non-zero.  The solution does not have a
simple expansion in integral powers of $R$ at the boundary, which makes
it more of a challenge for our spectral methods to have the accuracy
required to deal with the singularities in the equation.  Still, we
were able to approach very close to the trumpet limit, at least in the
spherically symmetric case, by using the analytic solution for
$\Omega/R$ at the boundary to formulate the boundary condition as a
Dirichlet condition on $\Omega$ at a small, but non-zero, $R$ and by
adding extra collocation points near the boundary.  If need be, reformulating the
Hamiltonian constraint as an equation for $\Omega/R$, as in
Eq.~(\ref{Eq:Trumpet2}), in a domain near the inner boundary and
explicitly canceling the singular terms at the boundary should make it
possible to manage exact trumpet boundary conditions numerically.

One interesting aspect of hyperboloidal Bowen-York data lies in the
physical interpretation of the spin parameter $S^i$ and the boost
parameter $P^i$. On asymptotically flat hypersurfaces (i.e.
$\TrKW=0$), one can evaluate the ADM integrals and find that the
Bowen-York parameter $P^i$ agrees with the ADM linear momentum, and
that $S^i$ agrees with the ADM angular momentum. For hyperboloidal
slices, the ADM formulas are not applicable. Nevertheless, a single
unboosted spinning black hole, because it is axisymmetric, has a
well-defined angular momentum which agrees with the spin parameter
$S^i$ (see Eq.~(\ref{eq:QuasilocalSpin})). The relationship of the
boost parameters to the linear momentum is less clear. The conformal
compactification leaves a rescaling freedom $R\to \Rrescale R$
unspecified (see Eq.~(\ref{eq:Rrescaling})), and as argued in
Sec.~\ref{SubSec:BhPlusSpinBoost}, the boost parameter rescales as
$P^i\to\Rrescale^{-1} P^i$, so that the vector $P^i$ {\em by itself}
has no physical meaning. However, one can define a scale invariant
quantity, $\Omega_{\rm max} P^i$, which may be considered a
``quasi-local'' linear momentum, at least when $\TrKW$ is small. The
proper interpretation of boosts on CMC hypersurfaces requires further
analysis.

When angular momentum is defined, we can consider the question of how
large spins can be constructed with hyperboloidal Bowen-York data. In
Sec.~\ref{Sec:NumResults:SingleSpin}, we have considered a sequence of
black hole initial data with increasing spins, and
Fig.~\ref{fig:MinMax2RicciScalar} shows that black holes have been
constructed with spin-extremality parameter\footnote{We avoid the more
  widely used spin measure $\chi=S/M^2$, with $M$ the Christoudoulou
  mass, because $\chi\le 1$ due to the definition of the
  Christoudoulou mass, and because the Christoudoulou mass only has
  physical meaning for Kerr black holes.} $\zeta=S/(2M_{\rm
  irr}^2)\approx 0.78$. In contrast, standard Bowen-York data for a
single spinning black hole allows $\zeta\lesssim 0.83$ (Fig.~2 of
Ref.~\cite{Lovelace2008}), whereas conformally flat conformal thin
sandwich data allows $\zeta\lesssim 0.56$ along the easily accessible
lower branch of solutions, and $\zeta\lesssim 0.87$ along the upper
branch (Fig.~7 of Ref.~\cite{Lovelace2008}). We thus see that
hyperboloidal Bowen-York initial data allows similarly large spins as
the standard Bowen-York initial data (this includes the widely used
puncture initial data~\cite{Brandt1997} as a special case). We have
not tested the sensitivity of the maximum achievable $\zeta$ to
variations of the other Bowen-York parameters $C$, $\TrKW$, $R_{\rm
  ms}$ of spinning black holes, but do not expect it to be large as
long as $K \Mirr$ is reasonably small.

% Harald's notes:  chi vs. zeta
% zeta   chi
% 0.50 0.8
% 0.52 0.818639798489
% 0.54 0.836172189532
% 0.56 0.852618757613
% 0.58 0.868003591739
% 0.60 0.882352941176
% 0.62 0.895694885871
% 0.64 0.908059023837
% 0.66 0.919476177208
% 0.68 0.929978118162
% 0.70 0.939597315436
% 0.72 0.948366701791
% 0.74 0.956319462393
% 0.76 0.963488843813
% 0.78 0.969907983089
% 0.80 0.975609756098
% 0.82 0.980626644343
% 0.84 0.984990619137
% 0.86 0.988733042079
% 0.88 0.991884580703
% 0.90 0.994475138122
% 0.92 0.996533795494
% 0.94 0.998088766192
% 0.96 0.999167360533
% 0.98 0.999795960008

The simplifying assumptions of Bowen-York initial data appear to limit
the ability to push towards near-extremal spins $\zeta\approx 1$.  To
construct larger spins, one would have to give up these simplifying
assumptions, most notably conformal flatness.  An approach based on
the extended conformal thin sandwich (XCTS) equations similar to
Ref.~\cite{Lovelace2008} seems very promising.  Note that for the
Schwarzschild spacetime, the {\em space-time} metric can be
conformally rescaled, resulting in conformal lapse and shift functions
which are finite at $\ScriPlus$ (see Eq.~(\ref{eq:Schwarz2})).  Thus,
it seems quite likely that the XCTS equations rewritten in suitably
rescaled variables can be used to construct more sophisticated
hyperboloidal initial data.  The XCTS--approach has another
interesting feature.  In this approach, the spins and boosts of the
black holes are implemented by {\em boundary conditions} at the black
hole horizons~\cite{Cook2002,Cook2004,Caudill-etal:2006}; the region
of the initial data hypersurface close to the black holes should only
be mildly affected by the ``warping up'' of the CMC hypersurface at
large radii as it approaches $\ScriPlus$.  Therefore, within the XCTS
framework, it might be easier to interpret a boost.  This will be a
topic of future research.

%%%%%%%%%%%%%%%%%%%%%%%%%%
\begin{acknowledgments}
%%%%%%%%%%%%%%%%%%%%%%%%%%
  We gratefully acknowledge Vincent Moncrief for originating this
  scheme and for numerous discussions since, and Richard Matzner,
  Oliver Rinne, and Olivier Sarbach for their many interactions and
  invaluable feedback. We also thank Geoffrey Lovelace for providing
  the code to compute the Ricci scalar on apparent horizons. The
  elliptic solver used here is part of the {\tt SpEC} code primarily
  developed by Lawrence Kidder, HP and Mark Scheel. LB and HP were
  supported in part by grants from the Sherman Fairchild Foundation
  and the Brinson Foundation, by NSF Grants No. PHY-0601459, No.
  PHY-0652995, and No. DMS-0553302. LB was also supported by grants
  NSF PHY 03 54842 and NASA NNG 04GL37G to the University of Texas at
  Austin.  HP gratefully acknowledges support from NSERC of Canada and
the Canadian Institute for Advanced Research.
\end{acknowledgments}

%%%%%%%%%%%%%%%%%%%%%%%%%%
\appendix*
\label{Appendix}
%%%%%%%%%%%%%%%%%%%%%%%%%%

\section{The Inversion Symmetry of the CMC Initial Value Problem}
In discussions of the conformally flat initial value problem, there
has been considerable interest in inversion symmetric initial value
data (see~\cite{Cook1993} and references therein). The issue of
inversion symmetry arises when the initial hypersurface contains a
minimal surface at a conformal radius $R=R_{\rm ms}$. As $R$ decreases
below this value, the physical radius $r$ increases and becomes
infinite in the limit $R \rightarrow 0$. The hypersurface may or may
not be symmetric under the inversion transformation $R \rightarrow
R_{\rm ms}^2/R$. Imposing inversion symmetry on the initial data, and
requiring that it be preserved during subsequent evolution, can lead
to relatively simple excision boundary conditions at $R=R_{\rm ms}$.
Inversion symmetry was discussed in the original Bowen and York paper
\cite{Bowen-York:1980}, and has been exploited in much of the
numerical work based on the Bowen-York class of solutions to the
initial value problem on maximal hypersurfaces. In this Appendix, we
show that the conditions on the solution for the conformal traceless
extrinsic curvature tensor which lead to inversion symmetry are the
same on CMC hypersurfaces as they are on maximal hypersurfaces,
noting, however, that the most general Bowen-York solution of the
conformal momentum constraint equation does not admit inversion
symmetry.

We start with the Hamiltonian constraint equation in the form given in
Eq.~(\ref{Eq:Trumpet2}) as an equation for the scale-invariant 
variable $U \equiv \Omega/R $.  A rearrangement of terms gives 
a form in which the possibility of inversion symmetry is manifest:
\begin{align}
  &R\frac{\partial } {{\partial R}}\left( {R\frac{{\partial U}}
      {{\partial R}}} \right) + \hat \Delta U + \frac{1}
  {2}U = \nonumber \\
  &\frac{3}{{2U}}\left[ {\left( {R\frac{{\partial U}} {{\partial R}}}
      \right)^2 + \hat \nabla U \cdot \hat \nabla U - \left( {\frac{\TrKW}
          {3}} \right)^2 + \frac{1} {6}U^6 W } \right],
\end{align}
where $\hat \Delta$ is the Laplacian operator and $\hat \nabla$ is the
gradient operator on the unit two-sphere. The only term not obviously
symmetric under the inversion transformation is the term involving the
source term $W$.  If and only if $W$ is inversion symmetric, $W\left(
  {R,\theta,\varphi}\right) = W\left( {R_{\rm ms}^2/R,\theta,\phi}
\right)$, will the solution for $U$, subject to the minimal surface
condition ${\partial U}/{\partial R} = 0$ at $R = R_{\rm ms}$, be
inversion symmetric, $U\left( {R,\theta,\varphi} \right) = U\left(
  {R_{\rm ms}^2/R,\theta,\phi} \right)$.

The generalized Bowen-York solution for $\tilde{\AW}_{ij} $ is given
in Eq.~(\ref{eq:GeneralMomentumTensor}). From this, we find
\begin{align}
\label{eq:AtildeSquared}
  W =&\; R^6 \tilde{\AW}_{ij} \tilde{\AW}^{ij} \nonumber \\
=&\; \frac{9} {2}R^2 \left[ {P^k P_k + 2\left( {P^i n_i } \right)\left(
        {P^j n_j } \right)} \right]\nonumber \\
  &+ \frac{9}{2}R^{-2} \left[ {Q^k Q_k + 2\left( {Q^i n_i } \right)\left(
       {Q^j n_j } \right)} \right]\nonumber \\
 & +6C^2 - 9\left[ {P^k Q_k  - 4\left( {P^i n_i } \right)\left( {Q^j n_j
        } \right)} \right]\nonumber \\
  &+ 18\left( {\varepsilon _{ijk} S^j n^k } \right)
  \left( {\varepsilon ^{imn} S_m n_n } \right)\nonumber \\
  &- 18C\left[ {R\left( {P^k n_k } \right) + \frac{1}
      {R}\left( {Q^k n_k } \right)} \right]\nonumber \\
  &- 18\left[ {R\left({\varepsilon _{ijk} P^i S^j n^k } \right) - \frac{1}
      {R}\left( {\varepsilon _{ijk} Q^i S^j n^k } \right)} \right]. 
\end{align}

Under an inversion transformation, the first two terms on the
right-hand side of Eq.~(\ref{eq:AtildeSquared}) transform into each
other provided that $Q^i = \pm R_{\rm ms}^2 P^i $. The next three
terms do not depend on $R$ and are therefore trivially
inversion-symmetric. Symmetry of the second to last square-bracket
requires $Q^i = + R_{\rm ms}^2 P^i $, while symmetry of the last
square-bracket requires $Q^i = - R_{\rm ms}^2 P^i $ unless it vanishes
because the boost and spin vectors are co-linear. Without any
restrictions on the Bowen-York parameters $\left( {C,P^i,S^i }
\right)$, there is no choice of the $Q^i$ which guarantees an
inversion-symmetric $W$ and therefore no guarantee of an
inversion-symmetric solution of the Hamiltonian constraint equation.
This result does not depend on the value of $\TrKW$. If we set $C=0$,
we recover the inversion symmetry result as usually stated for maximal
hypersurfaces, that the ``minus'' form of inversion symmetry applies
for general spin and boost.

%%%%%%%%%%%%%%%%%%%%%%%%%%
\bibliography{../References/References}

\begin{thebibliography}{62}
\expandafter\ifx\csname natexlab\endcsname\relax\def\natexlab#1{#1}\fi
\expandafter\ifx\csname bibnamefont\endcsname\relax
  \def\bibnamefont#1{#1}\fi
\expandafter\ifx\csname bibfnamefont\endcsname\relax
  \def\bibfnamefont#1{#1}\fi
\expandafter\ifx\csname citenamefont\endcsname\relax
  \def\citenamefont#1{#1}\fi
\expandafter\ifx\csname url\endcsname\relax
  \def\url#1{\texttt{#1}}\fi
\expandafter\ifx\csname urlprefix\endcsname\relax\def\urlprefix{URL }\fi
\providecommand{\bibinfo}[2]{#2}
\providecommand{\eprint}[2][]{\url{#2}}

\bibitem[{\citenamefont{Bondi et~al.}(1962)\citenamefont{Bondi, van~der Burg,
  and Metzner}}]{Bondi1962}
\bibinfo{author}{\bibfnamefont{H.}~\bibnamefont{Bondi}},
  \bibinfo{author}{\bibfnamefont{M.~G.~J.} \bibnamefont{van~der Burg}},
  \bibnamefont{and} \bibinfo{author}{\bibfnamefont{A.~W.~K.}
  \bibnamefont{Metzner}}, \bibinfo{journal}{Proc. R. Soc. Lond. A}
  \textbf{\bibinfo{volume}{269}}, \bibinfo{pages}{21} (\bibinfo{year}{1962}).

\bibitem[{\citenamefont{Sachs}(1962)}]{Sachs1962}
\bibinfo{author}{\bibfnamefont{R.~K.} \bibnamefont{Sachs}},
  \bibinfo{journal}{Proc. R. Soc. Lond. A} \textbf{\bibinfo{volume}{270}},
  \bibinfo{pages}{103} (\bibinfo{year}{1962}).

\bibitem[{\citenamefont{Sachs}(1964)}]{Sachs1964}
\bibinfo{author}{\bibfnamefont{R.~K.} \bibnamefont{Sachs}}, in
  \emph{\bibinfo{booktitle}{Relativity, Groups, and Topology}}, edited by
  \bibinfo{editor}{\bibfnamefont{C.~M.} \bibnamefont{DeWitt}} \bibnamefont{and}
  \bibinfo{editor}{\bibfnamefont{B.}~\bibnamefont{DeWitt}}
  (\bibinfo{publisher}{Gordon and Breach}, \bibinfo{address}{New York},
  \bibinfo{year}{1964}).

\bibitem[{\citenamefont{Stewart}(1989)}]{Stewart1989}
\bibinfo{author}{\bibfnamefont{J.~M.} \bibnamefont{Stewart}},
  \bibinfo{journal}{Proc. R. Soc. Lond. A} \textbf{\bibinfo{volume}{424}},
  \bibinfo{pages}{211} (\bibinfo{year}{1989}).

\bibitem[{\citenamefont{Nerozzi et~al.}(2005)\citenamefont{Nerozzi, Beetle,
  Bruni, Burko, and Pollney}}]{Nerozzi2005}
\bibinfo{author}{\bibfnamefont{A.}~\bibnamefont{Nerozzi}},
  \bibinfo{author}{\bibfnamefont{C.}~\bibnamefont{Beetle}},
  \bibinfo{author}{\bibfnamefont{M.}~\bibnamefont{Bruni}},
  \bibinfo{author}{\bibfnamefont{L.~M.} \bibnamefont{Burko}}, \bibnamefont{and}
  \bibinfo{author}{\bibfnamefont{D.}~\bibnamefont{Pollney}},
  \bibinfo{journal}{Phys.\ Rev.\ D} \textbf{\bibinfo{volume}{72}},
  \bibinfo{eid}{024014} (\bibinfo{year}{2005}).

\bibitem[{\citenamefont{Campanelli
  et~al.}(2006{\natexlab{a}})\citenamefont{Campanelli, Kelly, and
  Lousto}}]{Campanelli2006}
\bibinfo{author}{\bibfnamefont{M.}~\bibnamefont{Campanelli}},
  \bibinfo{author}{\bibfnamefont{B.}~\bibnamefont{Kelly}}, \bibnamefont{and}
  \bibinfo{author}{\bibfnamefont{C.~O.} \bibnamefont{Lousto}},
  \bibinfo{journal}{Phys.\ Rev.\ D} \textbf{\bibinfo{volume}{73}},
  \bibinfo{eid}{064005} (\bibinfo{year}{2006}{\natexlab{a}}).

\bibitem[{\citenamefont{Lehner and Moreschi}(2007)}]{Lehner2007}
\bibinfo{author}{\bibfnamefont{L.}~\bibnamefont{Lehner}} \bibnamefont{and}
  \bibinfo{author}{\bibfnamefont{O.~M.} \bibnamefont{Moreschi}},
  \bibinfo{journal}{Phys.\ Rev.\ D} \textbf{\bibinfo{volume}{76}},
  \bibinfo{eid}{124040} (\bibinfo{year}{2007}).

\bibitem[{\citenamefont{Calabrese et~al.}(2002)\citenamefont{Calabrese, Lehner,
  and Tiglio}}]{Calabrese2001}
\bibinfo{author}{\bibfnamefont{G.}~\bibnamefont{Calabrese}},
  \bibinfo{author}{\bibfnamefont{L.}~\bibnamefont{Lehner}}, \bibnamefont{and}
  \bibinfo{author}{\bibfnamefont{M.}~\bibnamefont{Tiglio}},
  \bibinfo{journal}{Phys.\ Rev.\ D} \textbf{\bibinfo{volume}{65}},
  \bibinfo{pages}{104031} (\bibinfo{year}{2002}).

\bibitem[{\citenamefont{Rinne}(2006)}]{Rinne2006}
\bibinfo{author}{\bibfnamefont{O.}~\bibnamefont{Rinne}},
  \bibinfo{journal}{Class.\ Quantum Grav.} \textbf{\bibinfo{volume}{23}},
  \bibinfo{pages}{6275} (\bibinfo{year}{2006}).

\bibitem[{\citenamefont{Buchman and Sarbach}(2006)}]{Buchman2006}
\bibinfo{author}{\bibfnamefont{L.~T.} \bibnamefont{Buchman}} \bibnamefont{and}
  \bibinfo{author}{\bibfnamefont{O.~C.~A.} \bibnamefont{Sarbach}},
  \bibinfo{journal}{Class.\ Quantum Grav.} \textbf{\bibinfo{volume}{23}},
  \bibinfo{pages}{6709} (\bibinfo{year}{2006}).

\bibitem[{\citenamefont{Rinne et~al.}(2007)\citenamefont{Rinne, Lindblom, and
  Scheel}}]{Rinne2007}
\bibinfo{author}{\bibfnamefont{O.}~\bibnamefont{Rinne}},
  \bibinfo{author}{\bibfnamefont{L.}~\bibnamefont{Lindblom}}, \bibnamefont{and}
  \bibinfo{author}{\bibfnamefont{M.~A.} \bibnamefont{Scheel}},
  \bibinfo{journal}{Class.\ Quantum Grav.} \textbf{\bibinfo{volume}{24}},
  \bibinfo{pages}{4053} (\bibinfo{year}{2007}).

\bibitem[{\citenamefont{Ruiz et~al.}(2007)\citenamefont{Ruiz, Rinne, and
  Sarbach}}]{Ruiz2007}
\bibinfo{author}{\bibfnamefont{M.}~\bibnamefont{Ruiz}},
  \bibinfo{author}{\bibfnamefont{O.}~\bibnamefont{Rinne}}, \bibnamefont{and}
  \bibinfo{author}{\bibfnamefont{O.}~\bibnamefont{Sarbach}},
  \bibinfo{journal}{Class.\ Quantum Grav.} \textbf{\bibinfo{volume}{24}},
  \bibinfo{pages}{6349} (\bibinfo{year}{2007}).

\bibitem[{\citenamefont{Rinne et~al.}(2009)\citenamefont{Rinne, Buchman,
  Scheel, and Pfeiffer}}]{Rinne2008b}
\bibinfo{author}{\bibfnamefont{O.}~\bibnamefont{Rinne}},
  \bibinfo{author}{\bibfnamefont{L.~T.} \bibnamefont{Buchman}},
  \bibinfo{author}{\bibfnamefont{M.~A.} \bibnamefont{Scheel}},
  \bibnamefont{and} \bibinfo{author}{\bibfnamefont{H.~P.}
  \bibnamefont{Pfeiffer}}, \bibinfo{journal}{Class.\ Quantum Grav.}
  \textbf{\bibinfo{volume}{26}}, \bibinfo{pages}{075009}
  (\bibinfo{year}{2009}).

\bibitem[{\citenamefont{Seiler et~al.}(2008)\citenamefont{Seiler, Szil\'agyi,
  Pollney, and Rezzolla}}]{Seiler2008}
\bibinfo{author}{\bibfnamefont{J.}~\bibnamefont{Seiler}},
  \bibinfo{author}{\bibfnamefont{B.}~\bibnamefont{Szil\'agyi}},
  \bibinfo{author}{\bibfnamefont{D.}~\bibnamefont{Pollney}}, \bibnamefont{and}
  \bibinfo{author}{\bibfnamefont{L.}~\bibnamefont{Rezzolla}},
  \bibinfo{journal}{Class.\ Quantum Grav.} \textbf{\bibinfo{volume}{25}},
  \bibinfo{pages}{175020} (\bibinfo{year}{2008}).

\bibitem[{\citenamefont{Winicour}(2009)}]{Winicour2009}
\bibinfo{author}{\bibfnamefont{J.}~\bibnamefont{Winicour}},
  \bibinfo{journal}{Living Rev.~Rel.} \textbf{\bibinfo{volume}{12}}
  (\bibinfo{year}{2009}),
  \urlprefix\url{http://www.livingreviews.org/lrr-2009-3}.

\bibitem[{\citenamefont{Stewart and Friedrich}(1982)}]{Stewart1982}
\bibinfo{author}{\bibfnamefont{J.~M.} \bibnamefont{Stewart}} \bibnamefont{and}
  \bibinfo{author}{\bibfnamefont{H.}~\bibnamefont{Friedrich}},
  \bibinfo{journal}{Proc. R. Soc. Lond. A} \textbf{\bibinfo{volume}{384}},
  \bibinfo{pages}{427} (\bibinfo{year}{1982}).

\bibitem[{\citenamefont{Friedrich}(1983)}]{Friedrich1983}
\bibinfo{author}{\bibfnamefont{H.}~\bibnamefont{Friedrich}},
  \bibinfo{journal}{Commun.\ Math.\ Phys.} \textbf{\bibinfo{volume}{91}},
  \bibinfo{pages}{445} (\bibinfo{year}{1983}).

\bibitem[{\citenamefont{Frauendiener}(2004)}]{FrauendienerLRR}
\bibinfo{author}{\bibfnamefont{J.}~\bibnamefont{Frauendiener}},
  \bibinfo{journal}{Living Rev.~Rel.} \textbf{\bibinfo{volume}{7}}
  (\bibinfo{year}{2004}),
  \urlprefix\url{http://www.livingreviews.org/lrr-2004-1}.

\bibitem[{\citenamefont{H{\"{u}}bner}(1999)}]{Huebner1999}
\bibinfo{author}{\bibfnamefont{P.}~\bibnamefont{H{\"{u}}bner}},
  \bibinfo{journal}{Class.\ Quantum Grav.} \textbf{\bibinfo{volume}{16}},
  \bibinfo{pages}{2823} (\bibinfo{year}{1999}).

\bibitem[{\citenamefont{Zengino\u{g}lu}(2008{\natexlab{a}})}]{Zenginoglu2008}
\bibinfo{author}{\bibfnamefont{A.}~\bibnamefont{Zengino\u{g}lu}},
  \bibinfo{journal}{Class.\ Quantum Grav.} \textbf{\bibinfo{volume}{25}},
  \bibinfo{pages}{195025} (\bibinfo{year}{2008}{\natexlab{a}}).

\bibitem[{\citenamefont{Moncrief and Rinne}(2009)}]{Moncrief2009}
\bibinfo{author}{\bibfnamefont{V.}~\bibnamefont{Moncrief}} \bibnamefont{and}
  \bibinfo{author}{\bibfnamefont{O.}~\bibnamefont{Rinne}},
  \bibinfo{journal}{Class.\ Quantum Grav.} \textbf{\bibinfo{volume}{26}},
  \bibinfo{pages}{125010} (\bibinfo{year}{2009}).

\bibitem[{\citenamefont{Bowen and {York, Jr.}}(1980)}]{Bowen-York:1980}
\bibinfo{author}{\bibfnamefont{J.~M.} \bibnamefont{Bowen}} \bibnamefont{and}
  \bibinfo{author}{\bibfnamefont{J.~W.} \bibnamefont{{York, Jr.}}},
  \bibinfo{journal}{Phys.\ Rev.\ D} \textbf{\bibinfo{volume}{21}},
  \bibinfo{pages}{2047} (\bibinfo{year}{1980}).

\bibitem[{\citenamefont{Brandt and Br{\"u}gmann}(1997)}]{Brandt1997}
\bibinfo{author}{\bibfnamefont{S.}~\bibnamefont{Brandt}} \bibnamefont{and}
  \bibinfo{author}{\bibfnamefont{B.}~\bibnamefont{Br{\"u}gmann}},
  \bibinfo{journal}{Phys.\ Rev.\ Lett.} \textbf{\bibinfo{volume}{78}},
  \bibinfo{pages}{3606} (\bibinfo{year}{1997}).

\bibitem[{\citenamefont{Pfeiffer et~al.}(2003)\citenamefont{Pfeiffer, Kidder,
  Scheel, and Teukolsky}}]{Pfeiffer2003}
\bibinfo{author}{\bibfnamefont{H.~P.} \bibnamefont{Pfeiffer}},
  \bibinfo{author}{\bibfnamefont{L.~E.} \bibnamefont{Kidder}},
  \bibinfo{author}{\bibfnamefont{M.~A.} \bibnamefont{Scheel}},
  \bibnamefont{and} \bibinfo{author}{\bibfnamefont{S.~A.}
  \bibnamefont{Teukolsky}}, \bibinfo{journal}{Comput.\ Phys.\ Commun.}
  \textbf{\bibinfo{volume}{152}}, \bibinfo{pages}{253} (\bibinfo{year}{2003}).

\bibitem[{\citenamefont{Zengino\u{g}lu}(2008{\natexlab{b}})}]{Zenginoglu2008b}
\bibinfo{author}{\bibfnamefont{A.}~\bibnamefont{Zengino\u{g}lu}},
  \bibinfo{journal}{Class.\ Quantum Grav.} \textbf{\bibinfo{volume}{25}},
  \bibinfo{pages}{145002} (\bibinfo{year}{2008}{\natexlab{b}}).

\bibitem[{\citenamefont{Ohme et~al.}(2009)\citenamefont{Ohme, Hannam, Husa, and
  {{\'O}~Murchadha}}}]{Hannam2009}
\bibinfo{author}{\bibfnamefont{F.}~\bibnamefont{Ohme}},
  \bibinfo{author}{\bibfnamefont{M.}~\bibnamefont{Hannam}},
  \bibinfo{author}{\bibfnamefont{S.}~\bibnamefont{Husa}}, \bibnamefont{and}
  \bibinfo{author}{\bibfnamefont{N.}~\bibnamefont{{{\'O}~Murchadha}}},
  \bibinfo{journal}{Class.\ Quantum Grav.} \textbf{\bibinfo{volume}{26}},
  \bibinfo{pages}{175014} (\bibinfo{year}{2009}).

\bibitem[{\citenamefont{Campanelli
  et~al.}(2006{\natexlab{b}})\citenamefont{Campanelli, Lousto, Marronetti, and
  Zlochower}}]{Campanelli2006a}
\bibinfo{author}{\bibfnamefont{M.}~\bibnamefont{Campanelli}},
  \bibinfo{author}{\bibfnamefont{C.~O.} \bibnamefont{Lousto}},
  \bibinfo{author}{\bibfnamefont{P.}~\bibnamefont{Marronetti}},
  \bibnamefont{and}
  \bibinfo{author}{\bibfnamefont{Y.}~\bibnamefont{Zlochower}},
  \bibinfo{journal}{Phys.\ Rev.\ Lett.} \textbf{\bibinfo{volume}{96}},
  \bibinfo{eid}{111101} (\bibinfo{year}{2006}{\natexlab{b}}).

\bibitem[{\citenamefont{Baker et~al.}(2006)\citenamefont{Baker, Centrella,
  Choi, Koppitz, and van Meter}}]{Baker2006a}
\bibinfo{author}{\bibfnamefont{J.~G.} \bibnamefont{Baker}},
  \bibinfo{author}{\bibfnamefont{J.}~\bibnamefont{Centrella}},
  \bibinfo{author}{\bibfnamefont{D.-I.} \bibnamefont{Choi}},
  \bibinfo{author}{\bibfnamefont{M.}~\bibnamefont{Koppitz}}, \bibnamefont{and}
  \bibinfo{author}{\bibfnamefont{J.}~\bibnamefont{van Meter}},
  \bibinfo{journal}{Phys.\ Rev.\ Lett.} \textbf{\bibinfo{volume}{96}},
  \bibinfo{eid}{111102} (\bibinfo{year}{2006}).

\bibitem[{\citenamefont{Hannam et~al.}(2007{\natexlab{a}})\citenamefont{Hannam,
  Husa, Pollney, Br{\"u}gmann, and {{\'O}~Murchadha}}}]{Hannam2006}
\bibinfo{author}{\bibfnamefont{M.}~\bibnamefont{Hannam}},
  \bibinfo{author}{\bibfnamefont{S.}~\bibnamefont{Husa}},
  \bibinfo{author}{\bibfnamefont{D.}~\bibnamefont{Pollney}},
  \bibinfo{author}{\bibfnamefont{B.}~\bibnamefont{Br{\"u}gmann}},
  \bibnamefont{and}
  \bibinfo{author}{\bibfnamefont{N.}~\bibnamefont{{{\'O}~Murchadha}}},
  \bibinfo{journal}{Phys.\ Rev.\ Lett.} \textbf{\bibinfo{volume}{99}},
  \bibinfo{pages}{241102} (\bibinfo{year}{2007}{\natexlab{a}}).

\bibitem[{\citenamefont{Hannam et~al.}(2007{\natexlab{b}})\citenamefont{Hannam,
  Husa, Br{\"u}gmann, Gonz{\'a}lez, Sperhake, and
  {{\'O}~Murchadha}}}]{Hannam2007d}
\bibinfo{author}{\bibfnamefont{M.}~\bibnamefont{Hannam}},
  \bibinfo{author}{\bibfnamefont{S.}~\bibnamefont{Husa}},
  \bibinfo{author}{\bibfnamefont{B.}~\bibnamefont{Br{\"u}gmann}},
  \bibinfo{author}{\bibfnamefont{J.~A.} \bibnamefont{Gonz{\'a}lez}},
  \bibinfo{author}{\bibfnamefont{U.}~\bibnamefont{Sperhake}}, \bibnamefont{and}
  \bibinfo{author}{\bibfnamefont{N.}~\bibnamefont{{{\'O}~Murchadha}}},
  \bibinfo{journal}{J. Phys.: Conf. Ser.} \textbf{\bibinfo{volume}{66}},
  \bibinfo{pages}{012047} (\bibinfo{year}{2007}{\natexlab{b}}).

\bibitem[{\citenamefont{Hannam et~al.}(2008)\citenamefont{Hannam, Husa, Ohme,
  Br{\"u}gmann, and {{\'O}~Murchadha}}}]{Hannam2008}
\bibinfo{author}{\bibfnamefont{M.}~\bibnamefont{Hannam}},
  \bibinfo{author}{\bibfnamefont{S.}~\bibnamefont{Husa}},
  \bibinfo{author}{\bibfnamefont{F.}~\bibnamefont{Ohme}},
  \bibinfo{author}{\bibfnamefont{B.}~\bibnamefont{Br{\"u}gmann}},
  \bibnamefont{and}
  \bibinfo{author}{\bibfnamefont{N.}~\bibnamefont{{{\'O}~Murchadha}}},
  \bibinfo{journal}{Phys.\ Rev.\ D} \textbf{\bibinfo{volume}{78}},
  \bibinfo{pages}{064020} (\bibinfo{year}{2008}).

\bibitem[{\citenamefont{Arnowitt et~al.}(1962)\citenamefont{Arnowitt, Deser,
  and Misner}}]{ADM}
\bibinfo{author}{\bibfnamefont{R.}~\bibnamefont{Arnowitt}},
  \bibinfo{author}{\bibfnamefont{S.}~\bibnamefont{Deser}}, \bibnamefont{and}
  \bibinfo{author}{\bibfnamefont{C.~W.} \bibnamefont{Misner}}, in
  \emph{\bibinfo{booktitle}{Gravitation: An Introduction to Current Research}},
  edited by \bibinfo{editor}{\bibfnamefont{L.}~\bibnamefont{Witten}}
  (\bibinfo{publisher}{Wiley}, \bibinfo{address}{New York},
  \bibinfo{year}{1962}).

\bibitem[{\citenamefont{{York, Jr.}}(1979)}]{york79}
\bibinfo{author}{\bibfnamefont{J.~W.} \bibnamefont{{York, Jr.}}}, in
  \emph{\bibinfo{booktitle}{Sources of Gravitational Radiation}}, edited by
  \bibinfo{editor}{\bibfnamefont{L.~L.} \bibnamefont{Smarr}}
  (\bibinfo{publisher}{Cambridge University Press},
  \bibinfo{address}{Cambridge, England}, \bibinfo{year}{1979}), pp.
  \bibinfo{pages}{83--126}.

\bibitem[{\citenamefont{Wald}(1984)}]{Wald}
\bibinfo{author}{\bibfnamefont{R.~M.} \bibnamefont{Wald}},
  \emph{\bibinfo{title}{General Relativity}} (\bibinfo{publisher}{University of
  Chicago Press}, \bibinfo{address}{Chicago and London}, \bibinfo{year}{1984}).

\bibitem[{\citenamefont{Misner et~al.}(1973)\citenamefont{Misner, Thorne, and
  Wheeler}}]{MTW}
\bibinfo{author}{\bibfnamefont{C.~W.} \bibnamefont{Misner}},
  \bibinfo{author}{\bibfnamefont{K.~S.} \bibnamefont{Thorne}},
  \bibnamefont{and} \bibinfo{author}{\bibfnamefont{J.~A.}
  \bibnamefont{Wheeler}}, \emph{\bibinfo{title}{Gravitation}}
  (\bibinfo{publisher}{Freeman}, \bibinfo{address}{New York, New York},
  \bibinfo{year}{1973}).

\bibitem[{\citenamefont{Cook and Pfeiffer}(2004)}]{Cook2004}
\bibinfo{author}{\bibfnamefont{G.~B.} \bibnamefont{Cook}} \bibnamefont{and}
  \bibinfo{author}{\bibfnamefont{H.~P.} \bibnamefont{Pfeiffer}},
  \bibinfo{journal}{Phys.\ Rev.\ D} \textbf{\bibinfo{volume}{70}},
  \bibinfo{pages}{104016} (\bibinfo{year}{2004}).

\bibitem[{\citenamefont{Penrose}(1965)}]{Penrose1965}
\bibinfo{author}{\bibfnamefont{R.}~\bibnamefont{Penrose}},
  \bibinfo{journal}{Proc. Roy. Soc. Lond. A} \textbf{\bibinfo{volume}{284}},
  \bibinfo{pages}{159} (\bibinfo{year}{1965}).

\bibitem[{\citenamefont{Murchadha and York{,
  Jr.}}(1974)}]{Murchadha-York:1974b}
\bibinfo{author}{\bibfnamefont{N.~{\'O}.} \bibnamefont{Murchadha}}
  \bibnamefont{and} \bibinfo{author}{\bibfnamefont{J.~W.} \bibnamefont{York{,
  Jr.}}}, \bibinfo{journal}{Phys.\ Rev.\ D} \textbf{\bibinfo{volume}{10}},
  \bibinfo{pages}{428} (\bibinfo{year}{1974}).

\bibitem[{\citenamefont{Pfeiffer and York}(2003)}]{Pfeiffer2003b}
\bibinfo{author}{\bibfnamefont{H.~P.} \bibnamefont{Pfeiffer}} \bibnamefont{and}
  \bibinfo{author}{\bibfnamefont{J.~W.} \bibnamefont{York}},
  \bibinfo{journal}{Phys.\ Rev.\ D} \textbf{\bibinfo{volume}{67}},
  \bibinfo{pages}{044022} (\bibinfo{year}{2003}).

\bibitem[{\citenamefont{York}(1973)}]{York1973}
\bibinfo{author}{\bibfnamefont{J.~W.} \bibnamefont{York}},
  \bibinfo{journal}{J.\ Math.\ Phys.} \textbf{\bibinfo{volume}{14}},
  \bibinfo{pages}{456} (\bibinfo{year}{1973}).

\bibitem[{\citenamefont{Brill et~al.}(1980)\citenamefont{Brill, Cavallo, and
  Isenberg}}]{Brill1980}
\bibinfo{author}{\bibfnamefont{D.~R.} \bibnamefont{Brill}},
  \bibinfo{author}{\bibfnamefont{J.~M.} \bibnamefont{Cavallo}},
  \bibnamefont{and} \bibinfo{author}{\bibfnamefont{J.~A.}
  \bibnamefont{Isenberg}}, \bibinfo{journal}{J.\ Math.\ Phys.}
  \textbf{\bibinfo{volume}{21}}, \bibinfo{pages}{2789} (\bibinfo{year}{1980}).

\bibitem[{\citenamefont{Malec and {{\'O}~Murchadha}}(2003)}]{Malec2003}
\bibinfo{author}{\bibfnamefont{E.}~\bibnamefont{Malec}} \bibnamefont{and}
  \bibinfo{author}{\bibfnamefont{N.}~\bibnamefont{{{\'O}~Murchadha}}},
  \bibinfo{journal}{Phys.\ Rev.\ D} \textbf{\bibinfo{volume}{68}},
  \bibinfo{pages}{124019} (\bibinfo{year}{2003}).

\bibitem[{\citenamefont{Malec and {{\'O}~Murchadha}}(2009)}]{Malec2009}
\bibinfo{author}{\bibfnamefont{E.}~\bibnamefont{Malec}} \bibnamefont{and}
  \bibinfo{author}{\bibfnamefont{N.}~\bibnamefont{{{\'O}~Murchadha}}},
  \bibinfo{journal}{Phys.\ Rev.\ D} \textbf{\bibinfo{volume}{80}},
  \bibinfo{pages}{024017} (\bibinfo{year}{2009}).

\bibitem[{\citenamefont{Bowen}(1979)}]{bowen79}
\bibinfo{author}{\bibfnamefont{J.~M.} \bibnamefont{Bowen}},
  \bibinfo{journal}{Gen.\ Relativ.\ Gravit.} \textbf{\bibinfo{volume}{11}},
  \bibinfo{pages}{227} (\bibinfo{year}{1979}).

\bibitem[{\citenamefont{Chru\'{s}ciel et~al.}(2004)\citenamefont{Chru\'{s}ciel,
  Jezierski, and Leski}}]{Chrusciel2004}
\bibinfo{author}{\bibfnamefont{P.~T.} \bibnamefont{Chru\'{s}ciel}},
  \bibinfo{author}{\bibfnamefont{J.}~\bibnamefont{Jezierski}},
  \bibnamefont{and} \bibinfo{author}{\bibfnamefont{S.}~\bibnamefont{Leski}},
  \bibinfo{journal}{Adv.\ Theor.\ Math.\ Phys.} \textbf{\bibinfo{volume}{8}},
  \bibinfo{pages}{83} (\bibinfo{year}{2004}).

\bibitem[{\citenamefont{Cook}(1990)}]{Thesis:Cook}
\bibinfo{author}{\bibfnamefont{G.~B.} \bibnamefont{Cook}}, Ph.D. thesis,
  \bibinfo{school}{University of North Carolina} (\bibinfo{year}{1990}).

\bibitem[{\citenamefont{Cook and {York, Jr.}}(1990)}]{cook90}
\bibinfo{author}{\bibfnamefont{G.~B.} \bibnamefont{Cook}} \bibnamefont{and}
  \bibinfo{author}{\bibfnamefont{J.~W.} \bibnamefont{{York, Jr.}}},
  \bibinfo{journal}{Phys.\ Rev.\ D} \textbf{\bibinfo{volume}{41}},
  \bibinfo{pages}{1077} (\bibinfo{year}{1990}).

\bibitem[{\citenamefont{Kulkarni et~al.}(1983)\citenamefont{Kulkarni, Shepley,
  and {York, Jr.}}}]{ksy83}
\bibinfo{author}{\bibfnamefont{A.~D.} \bibnamefont{Kulkarni}},
  \bibinfo{author}{\bibfnamefont{L.~C.} \bibnamefont{Shepley}},
  \bibnamefont{and} \bibinfo{author}{\bibfnamefont{J.~W.} \bibnamefont{{York,
  Jr.}}}, \bibinfo{journal}{Phys.\ Lett.} \textbf{\bibinfo{volume}{96A}},
  \bibinfo{pages}{228} (\bibinfo{year}{1983}).

\bibitem[{\citenamefont{Cook}(1991)}]{cook91}
\bibinfo{author}{\bibfnamefont{G.~B.} \bibnamefont{Cook}},
  \bibinfo{journal}{Phys.\ Rev.\ D} \textbf{\bibinfo{volume}{44}},
  \bibinfo{pages}{2983} (\bibinfo{year}{1991}).

\bibitem[{\citenamefont{Moncrief}(1986)}]{Moncrief86}
\bibinfo{author}{\bibfnamefont{V.}~\bibnamefont{Moncrief}},
  \bibinfo{journal}{Ann. Phys.} \textbf{\bibinfo{volume}{167}},
  \bibinfo{pages}{118} (\bibinfo{year}{1986}).

\bibitem[{\citenamefont{Saad}(1993)}]{Saad:1993}
\bibinfo{author}{\bibfnamefont{Y.}~\bibnamefont{Saad}}, \bibinfo{journal}{SIAM
  J. Sci. Comput} \textbf{\bibinfo{volume}{14}}, \bibinfo{pages}{461}
  (\bibinfo{year}{1993}).

\bibitem[{\citenamefont{Balay et~al.}(2009)\citenamefont{Balay, Buschelman,
  Gropp, Kaushik, Knepley, McInnes, Smith, and Zhang}}]{petsc-web-page}
\bibinfo{author}{\bibfnamefont{S.}~\bibnamefont{Balay}},
  \bibinfo{author}{\bibfnamefont{K.}~\bibnamefont{Buschelman}},
  \bibinfo{author}{\bibfnamefont{W.~D.} \bibnamefont{Gropp}},
  \bibinfo{author}{\bibfnamefont{D.}~\bibnamefont{Kaushik}},
  \bibinfo{author}{\bibfnamefont{M.~G.} \bibnamefont{Knepley}},
  \bibinfo{author}{\bibfnamefont{L.~C.} \bibnamefont{McInnes}},
  \bibinfo{author}{\bibfnamefont{B.~F.} \bibnamefont{Smith}}, \bibnamefont{and}
  \bibinfo{author}{\bibfnamefont{H.}~\bibnamefont{Zhang}},
  \emph{\bibinfo{title}{{PETSc} {W}eb {P}age}} (\bibinfo{year}{2009}),
  \bibinfo{note}{http://www.mcs.anl.gov/petsc}.

\bibitem[{\citenamefont{Balay et~al.}(2008)\citenamefont{Balay, Buschelman,
  Eijkhout, Gropp, Kaushik, Knepley, McInnes, Smith, and
  Zhang}}]{petsc-user-ref}
\bibinfo{author}{\bibfnamefont{S.}~\bibnamefont{Balay}},
  \bibinfo{author}{\bibfnamefont{K.}~\bibnamefont{Buschelman}},
  \bibinfo{author}{\bibfnamefont{V.}~\bibnamefont{Eijkhout}},
  \bibinfo{author}{\bibfnamefont{W.~D.} \bibnamefont{Gropp}},
  \bibinfo{author}{\bibfnamefont{D.}~\bibnamefont{Kaushik}},
  \bibinfo{author}{\bibfnamefont{M.~G.} \bibnamefont{Knepley}},
  \bibinfo{author}{\bibfnamefont{L.~C.} \bibnamefont{McInnes}},
  \bibinfo{author}{\bibfnamefont{B.~F.} \bibnamefont{Smith}}, \bibnamefont{and}
  \bibinfo{author}{\bibfnamefont{H.}~\bibnamefont{Zhang}}, \bibinfo{type}{Tech.
  Rep.} \bibinfo{number}{ANL-95/11 - Revision 3.0.0},
  \bibinfo{institution}{Argonne National Laboratory} (\bibinfo{year}{2008}).

\bibitem[{\citenamefont{Balay et~al.}(1997)\citenamefont{Balay, Gropp, McInnes,
  and Smith}}]{petsc_efficient}
\bibinfo{author}{\bibfnamefont{S.}~\bibnamefont{Balay}},
  \bibinfo{author}{\bibfnamefont{W.~D.} \bibnamefont{Gropp}},
  \bibinfo{author}{\bibfnamefont{L.~C.} \bibnamefont{McInnes}},
  \bibnamefont{and} \bibinfo{author}{\bibfnamefont{B.~F.} \bibnamefont{Smith}},
  in \emph{\bibinfo{booktitle}{Modern Software Tools in Scientific Computing}},
  edited by \bibinfo{editor}{\bibfnamefont{E.}~\bibnamefont{Arge}},
  \bibinfo{editor}{\bibfnamefont{A.~M.} \bibnamefont{Bruaset}},
  \bibnamefont{and} \bibinfo{editor}{\bibfnamefont{H.~P.}
  \bibnamefont{Langtangen}} (\bibinfo{publisher}{Birkh{\"a}user Press},
  \bibinfo{address}{Boston}, \bibinfo{year}{1997}), pp.
  \bibinfo{pages}{163--202}.

\bibitem[{\citenamefont{Pfeiffer et~al.}(2002)\citenamefont{Pfeiffer, Cook, and
  Teukolsky}}]{Pfeiffer2002a}
\bibinfo{author}{\bibfnamefont{H.~P.} \bibnamefont{Pfeiffer}},
  \bibinfo{author}{\bibfnamefont{G.~B.} \bibnamefont{Cook}}, \bibnamefont{and}
  \bibinfo{author}{\bibfnamefont{S.~A.} \bibnamefont{Teukolsky}},
  \bibinfo{journal}{Phys.\ Rev.\ D} \textbf{\bibinfo{volume}{66}},
  \bibinfo{pages}{024047} (\bibinfo{year}{2002}).

\bibitem[{\citenamefont{Pfeiffer et~al.}(2005)\citenamefont{Pfeiffer, Kidder,
  Scheel, and Shoemaker}}]{Pfeiffer2004}
\bibinfo{author}{\bibfnamefont{H.~P.} \bibnamefont{Pfeiffer}},
  \bibinfo{author}{\bibfnamefont{L.~E.} \bibnamefont{Kidder}},
  \bibinfo{author}{\bibfnamefont{M.~A.} \bibnamefont{Scheel}},
  \bibnamefont{and}
  \bibinfo{author}{\bibfnamefont{D.}~\bibnamefont{Shoemaker}},
  \bibinfo{journal}{Phys.\ Rev.\ D} \textbf{\bibinfo{volume}{71}},
  \bibinfo{pages}{024020} (\bibinfo{year}{2005}), \eprint{gr-qc/0410016}.

\bibitem[{\citenamefont{Dennison et~al.}(2006)\citenamefont{Dennison,
  Baumgarte, and Pfeiffer}}]{Dennison2006}
\bibinfo{author}{\bibfnamefont{K.~A.} \bibnamefont{Dennison}},
  \bibinfo{author}{\bibfnamefont{T.~W.} \bibnamefont{Baumgarte}},
  \bibnamefont{and} \bibinfo{author}{\bibfnamefont{H.~P.}
  \bibnamefont{Pfeiffer}}, \bibinfo{journal}{Phys.\ Rev.\ D}
  \textbf{\bibinfo{volume}{74}}, \bibinfo{pages}{064016}
  (\bibinfo{year}{2006}), \eprint{gr-qc/0606037}.

\bibitem[{\citenamefont{Lovelace et~al.}(2008)\citenamefont{Lovelace, Owen,
  Pfeiffer, and Chu}}]{Lovelace2008}
\bibinfo{author}{\bibfnamefont{G.}~\bibnamefont{Lovelace}},
  \bibinfo{author}{\bibfnamefont{R.}~\bibnamefont{Owen}},
  \bibinfo{author}{\bibfnamefont{H.~P.} \bibnamefont{Pfeiffer}},
  \bibnamefont{and} \bibinfo{author}{\bibfnamefont{T.}~\bibnamefont{Chu}},
  \bibinfo{journal}{Phys.\ Rev.\ D} \textbf{\bibinfo{volume}{78}},
  \bibinfo{pages}{084017} (\bibinfo{year}{2008}).

\bibitem[{\citenamefont{Rinne}(2009)}]{Rinne2009}
\bibinfo{author}{\bibfnamefont{O.}~\bibnamefont{Rinne}},
  \emph{\bibinfo{title}{An axisymmetric evolution code for the {E}instein
  equations on hyperboloidal slices}} (\bibinfo{year}{2009}),
  \eprint{arXiv:0910.0139v1[gr-qc]}.

\bibitem[{\citenamefont{Cook}(2002)}]{Cook2002}
\bibinfo{author}{\bibfnamefont{G.~B.} \bibnamefont{Cook}},
  \bibinfo{journal}{Phys.\ Rev.\ D} \textbf{\bibinfo{volume}{65}},
  \bibinfo{pages}{084003} (\bibinfo{year}{2002}).

\bibitem[{\citenamefont{{Caudill} et~al.}(2006)\citenamefont{{Caudill}, {Cook},
  {Grigsby}, and {Pfeiffer}}}]{Caudill-etal:2006}
\bibinfo{author}{\bibfnamefont{M.}~\bibnamefont{{Caudill}}},
  \bibinfo{author}{\bibfnamefont{G.~B.} \bibnamefont{{Cook}}},
  \bibinfo{author}{\bibfnamefont{J.~D.} \bibnamefont{{Grigsby}}},
  \bibnamefont{and} \bibinfo{author}{\bibfnamefont{H.~P.}
  \bibnamefont{{Pfeiffer}}}, \bibinfo{journal}{Phys.\ Rev.\ D}
  \textbf{\bibinfo{volume}{74}}, \bibinfo{pages}{064011}
  (\bibinfo{year}{2006}).

\bibitem[{\citenamefont{Cook et~al.}(1993)\citenamefont{Cook, Choptuik, Dubal,
  Klasky, Matzner, and Oliveira}}]{Cook1993}
\bibinfo{author}{\bibfnamefont{G.~B.} \bibnamefont{Cook}},
  \bibinfo{author}{\bibfnamefont{M.~W.} \bibnamefont{Choptuik}},
  \bibinfo{author}{\bibfnamefont{M.~R.} \bibnamefont{Dubal}},
  \bibinfo{author}{\bibfnamefont{S.}~\bibnamefont{Klasky}},
  \bibinfo{author}{\bibfnamefont{R.~A.} \bibnamefont{Matzner}},
  \bibnamefont{and} \bibinfo{author}{\bibfnamefont{S.~R.}
  \bibnamefont{Oliveira}}, \bibinfo{journal}{Phys.\ Rev.\ D}
  \textbf{\bibinfo{volume}{47}}, \bibinfo{pages}{1471} (\bibinfo{year}{1993}).

\end{thebibliography}
%%%%%%%%%%%%%%%%%%%%%%%%%%

\end{document}